\documentclass[manuscript,screen]{acmart}
\usepackage{bm}
\usepackage{multirow}
\usepackage{ulem}
\usepackage{bbding}
\usepackage{subfig}
\usepackage{enumitem}
\usepackage{xcolor}
\usepackage{threeparttable}
\AtBeginDocument{%
  }

\setcopyright{acmcopyright}
\copyrightyear{2024}
\acmYear{2024}
\acmDOI{XXXXXXX.XXXXXXX}

\acmJournal{JACM}
\acmVolume{1}
\acmNumber{1}
\acmArticle{1}
\acmMonth{1}

\begin{document}

\title{Disentangled  Cascaded Graph Convolution Networks 
for Multi-Behavior Recommendation}


%\author{Zhiyong Cheng}
%\affiliation{%
%  \institution{Key Laboratory of Knowledge Engineering with Big Data, Hefei University of Technology
%and Institute of Artificial Intelligence, Hefei Comprehensive National Science Centery}
%    \streetaddress{No. 485, Danxia Road}
%  \city{Hefei}
%  \state{Anhui}
%  \country{China}
%  \postcode{230009}
%}
%\email{jason.zy.cheng@gmail.com}

\author{Zhiyong Cheng}
\affiliation{%
  \institution{School of Computer Science and Information Engineering, Hefei University of Technology}
    \streetaddress{No. 485, Danxia Road}
  \city{Hefei}
  \state{Anhui}
  \country{China}
  \postcode{230009}
}
\email{jason.zy.cheng@gmail.com}

\author{Jianhua Dong}
\affiliation{%
  \institution{Shandong Artificial Intelligence Institute, Qilu University of Technology (Shandong Academy of Sciences)}
    \streetaddress{No. 19, Keyuan Road}
  \city{Jinan}
  \state{Shandong}
  \country{China}
  \postcode{250014}
}
\email{aahua.jh.dong@outlook.com}

\author{Fan Liu}
\affiliation{%
  \institution{School of Computing, National University of Singapore}
    \streetaddress{21 Lower Kent Ridge Road}
  \city{Singapore}
  \country{Singapore}
  \postcode{119077}
}
\email{liufancs@gmail.com}

\author{Lei Zhu}
\affiliation{%
  \institution{School of Electronic and Information Engineering, University of Tongji}
  \streetaddress{No. 4800 Caoan Road}
  \city{Shanghai}
  \country{China}
  \postcode{201804}
}

\email{leizhu0608@gmail.com}

\author{Xun Yang}
\affiliation{%
  \institution{School of Information Science and Technology, University of Science and Technology of China}
  \streetaddress{No. 443 Huangshan Road}
  \city{Hefei}
  \state{Anhui}
  \country{China}
  \postcode{230027}
}

\email{xyang21@ustc.edu.cn}

\author{Meng Wang}
\affiliation{%
  \institution{Key Laboratory of Knowledge Engineering with Big Data, Hefei University of Technology
and Institute of Artificial Intelligence, Hefei Comprehensive National Science Centery}
    \streetaddress{No. 485, Danxia Road}
  \city{Hefei}
  \state{Anhui}
  \country{China}
  \postcode{230009}
}
\email{eric.mengwang@gmail.com}

% \affiliation{%

% \affiliation{%

\renewcommand{\shortauthors}{Zhiyong Cheng, Jianhua Dong, Fan Liu, Lei Zhu, Xun Yang and Meng Wang.}

\begin{abstract}

Multi-behavioral recommender systems have emerged as a solution to address data sparsity and cold-start issues by incorporating auxiliary behaviors alongside target behaviors. However, existing models struggle to accurately capture varying user preferences across different behaviors and fail to account for diverse item preferences within behaviors. Various user preference factors (such as price or quality) entangled in the behavior may lead to sub-optimization problems. Furthermore, these models overlook the personalized nature of user behavioral preferences by employing uniform transformation networks for all users and items. To tackle these challenges, we propose the Disentangled Cascaded Graph Convolutional Network (Disen-CGCN), a novel multi-behavior recommendation model. Disen-CGCN employs disentangled representation techniques to effectively separate factors within user and item representations, ensuring their independence. In addition, it incorporates a multi-behavioral meta-network, enabling personalized feature transformation across user and item behaviors. Furthermore, an attention mechanism captures user preferences for different item factors within each behavior. By leveraging attention weights, we aggregate user and item embeddings separately for each behavior, computing preference scores that predict overall user preferences for items.  Our evaluation on benchmark datasets demonstrates the superiority of Disen-CGCN over state-of-the-art models, showcasing an average performance improvement of 7.07\% and 9.00\% on respective datasets. These results highlight Disen-CGCN's ability to effectively leverage multi-behavioral data, leading to more accurate recommendations.

\end{abstract}


\begin{CCSXML}
<ccs2012>
    <concept>
        <concept_id>10002951.10003317.10003331.10003271</concept_id>
        <concept_desc>Information systems~Personalization</concept_desc>
        <concept_significance>500</concept_significance>
        </concept>
        <concept>
        <concept_id>10002951.10003317.10003347.10003350</concept_id>
        <concept_desc>Information systems~Recommender systems</concept_desc>
        <concept_significance>500</concept_significance>
        </concept>
        <concept>
        <concept_id>10002951.10003227.10003351.10003269</concept_id>
        <concept_desc>Information systems~Collaborative filtering</concept_desc>
        <concept_significance>500</concept_significance>
    </concept>
</ccs2012>
\end{CCSXML}
\ccsdesc[500]{Information systems~Personalization}
\ccsdesc[500]{Information systems~Recommender systems}
\ccsdesc[500]{Information systems~Collaborative filtering}

\keywords{Multi-Behavior Recommendation, disentangled representation learning,  graph convolutional network, multi-behavior recommendation,  multi-task learning}


\maketitle

\section{Introduction}
%In the age of information explosion, we are exposed to various information in our daily life. It is crucial to find information that is useful to us quickly and efficiently. Recommender system is an effective solution\cite{he2020lightgcn,he2017neural,koren2009matrix} to perform information filtering and retrieval in a large amount of information. It has also been one of the core technologies of various platforms, like social networking platforms, e-commerce and news websites. Among the wide variety of recommendation methods, collaborative filtering(CF) techniques\cite{herlocker1999algorithmic,sarwar2001item,liu2021interest,cheng2022feature} have achieved great success and are widely used in academia and industry. The CF methods are used to learn user and item representations by modeling user-item interaction data. Following this methodology, recommender systems have evolved from early shallow models\cite{cheng2018aspect,rendle2012bpr,koren2009matrix} to deep models\cite{he2017neural,he2017nFM,cheng2016wide} utilizing deep learning, where deep neural network-based models (DNNs) can better capture the complex relationships between users and items. With the rise of Graph Convolutional Networks (GCNs), GCNs utilize higher-order neighbor aggregation of user-item interaction graphs to capture higher-order information about users and items\cite{wang2019neural,liu2020attribute,he2020lightgcn}. 
In the information era, we encounter an immense variety of information in our daily lives.  Efficiently navigating this overwhelming amount of information to quickly find what is relevant and useful is becoming increasingly crucial. In this context, recommender systems have become an indispensable tool, aiding in the filtering and retrieval of desired information from this abundance~\cite{he2020lightgcn,he2017neural,koren2009matrix}. These systems have established themselves as foundational technologies across diverse platforms, ranging from social networking sites to e-commerce platforms and news portals. Among various recommendation strategies, collaborative filtering (CF)~\cite{koren2009matrix,sarwar2001item,cheng2022feature} stands out as a particularly effective method. CF methods focus on learning user and item representations by analyzing user-item interaction data, evolving from basic, shallow models~\cite{cheng2018aspect,rendle2012bpr,koren2009matrix} to more complex deep models~\cite{he2017neural,he2017nFM,cheng2016wide}. The introduction of Graph Convolutional Networks (GCNs) has further advanced these systems, offering superior capabilities in modeling complex relationships and capturing detailed, higher-order information in user-item interaction graphs, thus significantly enhancing the precision and effectiveness of recommender systems~\cite{he2020lightgcn,liu2021interest,gao2023tors}.

Despite their efficacy,  current CF models often focus narrowly on single, profit-driven behaviors, neglecting the multi-dimensional nature of user interactions such as \textit{browsing}, \textit{adding to cart}, and \textit{purchasing}. These multi-behavior interactions offer a more holistic view of user preferences and can mitigate issues like data sparsity and cold-start problems in recommender systems. Recognizing this, recent research has shifted towards multi-behavior recommendation models, which consider a range of user interactions to enhance recommendation accuracy~\cite{gao2019learning,jin2020multi,qiu2018bprh,schlichtkrull2018modeling,xia2020multiplex,xia2021multi,yan2023cascading,cheng2023multi}.  
Recent developments in multi-behavior recommendation models have spurred notable innovations in the field. Typically, these models categorize behaviors into two types: \textit{target behaviors} (e.g., ``purchase"), which are directly linked to platform profits, and \textit{auxiliary behaviors} (e.g., ``browsing"), which provide rich supplementary data. This data from auxiliary behaviors is then leveraged to enhance recommendations for target behaviors. Early methods in this area extended standard matrix factorization into multiple matrices, enabling the direct modeling of multi-behavior interactions~\cite{krohn2012multi,singh2008relational,loni2016bayesian}. The introduction of deep learning and Graph Convolutional Networks (GCNs) has further refined these techniques. For example, the MATN model~\cite{xia2020MATN} incorporates transformer-based networks and memory-attention mechanisms to differentiate user-item relationships. Simultaneously, GNMR~\cite{xia2021multi} utilizes a relational aggregation network, capturing inter-behavior dependencies through recursive embedding propagation within a unified multi-behavior interaction graph. Notably, recent works have shown that recognizing the sequential nature of behaviors can lead to cutting-edge performance in recommendation systems~\cite{meng2023parallel,yan2023cascading,cheng2023multi}. For instance, MB-CGCN~\cite{cheng2023multi} effectively models behavioral dependencies in a sequential chain, utilizing user and item embeddings learned from one behavior as input features for the subsequent behavior's embedding learning process. Concurrently, MB-CGCN acknowledges that information transfer between behaviors may introduce noise or irrelevant information. To address this, a feature transformation module is strategically implemented prior to each behavior's information transfer phase..

While significant progress have been made in multi-behavior recommender systems, existing methods fall short in capturing the nuanced preferences users exhibit towards items across different behaviors. For instance, a user might browse a T-shirt on an e-commerce platform, drawn by its color and style, add it to the cart due to its affordability, and finally make the purchase, influenced by both price and perceived value. Capturing these shifting preferences at each stage is crucial for more accurate and personalized recommendations. Additionally, in the recent advanced cascading multi-behavior recommendation models, like CRGCN~\cite{yan2023cascading} or MB-CGCN~\cite{cheng2023multi}, when  transferring information from one behavior to a later one, they typically use a shared feature transformation function. However, preference patterns can vary significantly among individuals across different behaviors. A common transformation may not effectively personalize this process for individual users and items.

To address these gaps, we introduce the Disentangled Cascaded Graph Convolutional Network (Disen-CGCN), a sophisticated model that offers fine-grained analysis of multi-behavior user preferences. Disen-CGCN utilizes LightGCN as its foundational network,  chosen for its simplicity and effectiveness in learning user and item embeddings in a bipartite graph. Inspired by the success of the previous models~\cite{yan2023cascading,cheng2023multi}, our model also exploits the cascade relationship between behaviors, ensuring that the insights gained from one behavior inform the next. Distinguish from them,
% within each behavior, we employ disentanglement representation techniques to separate the various factors entangled in the user and item embeddings. Our model further incorporates a unique meta-network, facilitating personalized feature transformation between behaviors, thereby enhancing the accuracy of the recommendations. Additionally, we design an attention mechanism to capture the nuances of users' varying preferences for different factors across behaviors.  
we employ disentangled representation techniques in behavior, which is the basis for modeling fine-grained preferences, working to separate different factors of user and item representations (e.g., color, price, etc.) and ensuring that the different factors are independent of each other. Based on this foundation, our model further incorporates a unique meta-network to extract personalized information about users and items in each behavior to perform feature preference transfer between behaviors. Thus, the module comes to model fine-grained preference transfer between different behaviors of users and items. In addition, we design an attention mechanism that works to model the degree of fine-grained attention users pay to different factors of an item in each behavior to capture the nuances of users' different preferences for different factors in different behaviors.
The attention weights derived from this mechanism are subsequently utilized in a linear aggregation of user embeddings for the final prediction.  To evaluate the effectiveness of our proposed Disen-CGCN model, we conducted extensive experiments on  two benchmark datasets. The results show that the Disen-CGCN model significantly outperforms both single-behavior and recent advanced multi-behavior recommendation methods, achieving an average performance improvement of 7.07\% and 9.00\%, respectively. We also conducted detailed analyses to understand user preferences in different behaviors and assessed the impact of key model components and hyperparameters.

In summary, the main contributions are as follows:
\begin{itemize}
    \item We highlight the critical need to capture users' varying preferences across different behaviors in multi-behavior recommendation systems, and how personalized transformations between these behaviors can enhance the quality of user and item representations.
    
    \item We introduce Disen-CGCN, a model that harnesses the cascade relationship of multiple behaviors. It uses disentangled representation to clarify user and item preferences within behaviors and implements a meta-network for personalized feature transformation between behaviors.
    
    \item Our empirical results validate Disen-CGCN’s superior performance over benchmark models on real-world datasets, confirming its effectiveness in delivering nuanced and user-centric recommendations. We release our code for reproducibility\footnote{https://github.com/JianhuaDongCS/Disen-CGCN}.
\end{itemize}

The rest of this paper is organized as follows: Section~\ref{Related Work} provides an overview of the related work. Section~\ref{methodology} delves into the details of our Disen-CGCN model. Following this, Section~\ref{experiment} presents the experimental setup and discusses the results obtained from these experiments. The paper concludes with Section~\ref{conclusion}, summarizing the key findings and contributions.
\section{RELATED WORK} \label{Related Work}
\subsection{Collaborative Filtering}
Collaborative Filtering (CF)~\cite{hu2008collaborative,koren2010factor,zhou2023selfcf} has garnered significant attention in both academic and industrial communities as a pivotal approach in the development of recommender systems. CF models leverage user-item interaction data to learn their representations, typically embodied as embedding vectors in a shared latent space. These vectors are then used via interaction functions to predict user preferences for items.  Matrix Factorization (MF)~\cite{koren2009matrix} is a prime example of CF, known for its simplicity and efficacy, notably in the Netflix competition. Various MF variants, such as WRMF~\cite{hu2008collaborative}, BPR~\cite{rendle2012bpr}, PMF~\cite{mnih2007probabilistic}, and NMF~\cite{lee1999learning}, have been developed.  With the advent of deep learning, CF techniques have undergone significant advancement. Deep learning's robust expressive capabilities enable the learning of more sophisticated user and item embeddings and the capture of complex interactions between users and items~\cite{xue2017deep,he2018nais,cheng2018aspect,wu2016collaborative,mao2023finalmlp}. For example, NeuMF~\cite{he2017neural} combines a generalized factorization model with a deep multilayer perceptron to intricately model these interactions. FinalMLP~\cite{mao2023finalmlp} effectively harnesses a two-stream MLP model, demonstrating that the straightforward combination of two MLPs can yield unexpectedly strong performance. This model is further enhanced with a feature selection layer and an interaction aggregation layer, leading to substantial performance improvements. 

Recently, Graph Convolutional Networks (GCNs) have brought a paradigm shift in CF, modeling higher-order connections between users and items~\cite{wang2019neural,zhang2023tors,he2020lightgcn,liu2021interest,mao2021ultragcn,fan2022graph}.  LightGCN~\cite{he2020lightgcn} stands out as a simplified and effective GCN model by removing  the transformation matrix and nonlinear activation function, focusing exclusively on neighbor aggregation.  UltraGCN~\cite{mao2021ultragcn} advances this concept further by bypassing explicit message passing, effectively simulating the effect of infinite message passing layers. GTN~\cite{fan2022graph} introduces an innovative graph trend collaborative filtering approach, proposing a new graph trend filtering network to adaptively capture the reliability of user-item interactions.  The recent integration of sophisticated techniques like disentangled learning and self-supervised learning into GCN models further marks a further advancement in recommender systems.  This fusion has led to the creation of more powerful GCN-based recommendation models, as exemplified by ~\cite{wang2020disentangled,tao2022self,cai2023lightgcl,xia2022hypergraph,liu2021interest}. These innovative approaches have considerably enhanced the capabilities of recommendation systems, highlighting the ongoing evolution and adaptability of GCN applications in this field. However, despite these improvements, a common limitation persists: these models primarily rely on single-behavior user-item interaction data. This reliance often results in data sparsity, which poses a challenge to the effectiveness and accuracy of the recommendation process.

\subsection{Multi-behavior Recommendations}

Multi-behavior recommendation, which utilizes diverse behavioral data from user-item interactions, significantly enhances recommendation performance by addressing data sparsity and cold-start issues in recommender systems~\cite{xia2021multi,yan2023cascading,cheng2023multi,xin2023improving,meng2023coarse,xu2023multi,yan2023mb}. Early approaches in this domain, such as CMF~\cite{singh2008relational} and BF~\cite{zhao2015improving}, extended traditional matrix decomposition techniques to accommodate multiple matrices and shared embeddings across different behaviors. The incorporation of deep learning has further advanced multi-behavior recommendation systems. For instance,  NMTR~\cite{gao2019learning} models multi-behavior data by capturing the cascading relationships among various user behaviors, employing a multi-task learning approach for comprehensive training. Similarly, MBGCN~\cite{jin2020multi} learns user preferences via a unified interaction graph, enhancing embeddings through item-to-item propagation and integrating the impacts of various behaviors on final predictions.   MBGMN~\cite{xia2021graph} employs a meta graph neural network to model multi-behavior data, effectively learning the diversity and heterogeneity of interactions. More recent advancements include the use of contrastive learning ~\cite{wei2022contrastive,xu2023multi,gu2022self}. An example is MBSSL~\cite{xu2023multi}, which utilizes a behavior-aware GNN with self-attention to capture the subtleties and dependencies among behaviors, coupled with self-supervised learning for enhanced node differentiation.  Additionally, state-of-the-art methods now focus on modeling behaviors in a sequential chain~\cite{yan2023cascading,cheng2023multi,meng2023parallel}.  This approach utilizes embeddings learned from one behavior as input for learning embeddings in subsequent behaviors, thereby capturing the intricate dependencies within multi-behavior interactions.  A notable example is MB-CGCN~\cite{cheng2023multi}, which effectively utilizes cascading relationships between behaviors. It includes a feature transformation module designed to reduce noise and prevent the transfer of misleading information, further refining the recommendation process.

In this study, we introduce the Disentangled Cascaded Graph Convolutional Network (Disen-CGCN), a multi-behavior recommendation model that operates at a more granular level. This model represents an advancement over previous multi-behavior recommendation approaches by capturing more detailed user preferences across behaviors and managing information transfer between them, leading to personalized feature transformation. This granular approach offers deeper insights into individual user preferences within a multi-behavior framework.
%In this work, we design a multi-behavior recommendation model modeled at a finer granularity, the Disentangled Cascaded Graph Convolutional Network (Disen-CGCN). Compared to previous work on multi-behavior recommendation, our advantage is that we can capture more fine-grained user preferences for items in different behaviors as well as the information between behaviors to achieve personalized feature transformation.

\subsection{Disentangled Representation Learning}

Disentangled representation learning (DRL) aims to isolate and identify key explanatory factors in data, and in particular, DRL has attracted significant interest in image and text representation learning ~\cite{john2018disentangled,higgins2016beta}. For example, beta-VAE~\cite{higgins2016beta} utilizes a variational autoencoder framework aimed at unsupervised discovery of interpretable decomposition potential representations from raw image data.

In recommender systems, DRL has made significant progress by efficiently decomposing users' preferences for various factors~\cite{ma2019learning,wang2020disentangled,liu2022knowledge,liu2022disentangled,li2022disentangled,ren2023disentangled}. Early DRL-based recommendation models typically used variational autoencoders to transform user preferences into distributions, and therefore utilized the bootstrap of KL divergence to constrain the independence of the latent representations in order to penalize the discrepancy between the representations in the latent space and the prior~\cite{mathieu2019disentangling}. For example, MacridVAE~\cite{ma2019learning} distinguishes between different user intentions by analyzing them from both macro and micro perspectives. ADDVAE~\cite{tran2022aligning} generates doubly segregated representations from user-item interactions and textual content, and then aligns them to provide a more comprehensive understanding of user preferences and behaviors. To better understand the various intentions behind user-item interactions, user preferences are converted into embeddings, and the introduction of distance correlation~\cite{szekely2007measuring,szekely2009brownian} also enables the independence of any two pairs of embeddings. DGCF~\cite{wang2020disentangled} models the distributions of these intentions and progressively refines the intention-aware interaction graph and its representation. KMBFD~\cite {liu2022knowledge} employs a factor entanglement approach to derive multidimensional representations of users and items, integrating complex user-item interaction graphs with knowledge graph features. To ensure the robustness of these representations, it uses dimension classifiers specifically designed to ensure the independence of these decomposed dimensions. DMRL~\cite{liu2022disentangled} employs a strategy for decomposing factors in each modality in multimodal recommendation, which incorporates a multimodal attention mechanism to accurately capture the user's preference for each factor for the different modalities. Disen-GNN~\cite{li2022disentangled} distinguishes between different factors of user intent in session-based recommendation. It effectively reveals the user's specific purpose in a target session by integrating decentralized representation learning in Gated Graph Neural Networks (GGNNs). Recently, methods based on contrastive learning have also been attempted to be applied to disentangled representations, which are implemented to bring the same factors closer and push different factors further away through contrast loss. For example, DCCF~\cite{ren2023disentangled} is a disentangled contrast collaborative filtering framework that achieves intent disentanglement in an adaptive manner by utilizing contrast learning. AD-DRL~\cite{li2023attribute} utilizes contrastive learning to disentangle factors at the attribute and attribute-value level by assigning specific attributes to each of the factors in a multimodal feature.

While disentangled representation techniques have proven to be very successful in recommender systems, their application to multi-behavioral recommendation tasks remains limited. This study fills the gap in this area by introducing a disentangled cascade graph neural network model designed specifically for multi-behavioral recommendation.

\section{METHODOLOGY} \label{methodology}
\subsection{Problem Formulation}
%In real-world scenarios of online recommender systems, users tend to have multiple types of interactions in the platform, such as browsing, adding to cart and buying, etc. However, platforms, in order to gain more profit (e.g., buy), usually design traditional recommendation models to utilize only a single type of user-item interaction (a.k.a., target behavior). Such paradigms tend to achieve poor performance as they suffer from severe data sparsity and cold-start problems. Auxiliary behaviors in the platform (e.g., browsing and adding to cart, etc.) will contain rich interest tendencies of the user, and utilizing auxiliary behaviors to augment the information of the target behavior will usually alleviate the above problems and also improve the performance of the recommender system. In this work, we propose a multi-behavior recommendation model that models a finer-grained approach to exploring different levels of user preferences in each behavior. Before formally presenting our proposed Disen-CGCN, we first introduce the key notation and problem description.

In the context of online recommender systems, users typically engage in a variety of interactions on platforms, such as browsing, adding items to cart, and making purchases. Traditional recommendation models, often designed to maximize profit, usually focus on a single type of user-item interaction (known as target behavior). This approach, however, tends to be less effective due to significant data sparsity and cold-start issues. Auxiliary behaviors on the platform, like browsing and adding items to cart, provide valuable insights into users' interest patterns. Leveraging these auxiliary behaviors to enrich the information of the target behavior can mitigate these issues and enhance the overall performance of the recommender system. In this work, we introduce a multi-behavior recommendation model that adopts a more nuanced approach to understand varying levels of user preferences associated with each behavior. Prior to delving into our proposed Disen-CGCN model, we will first define key terms and outline the problem we aim to address.

Let $\mathcal{U}=(u_{1},u_{2},\dots ,u_{M} )$ and $\mathcal{I}=(i_{1},i_{2},\dots ,i_{N} )$ represent the set of users and items respectively, where $M$ and $N$ are the total number of users and items. We define $\mathbf{G}=(\mathbf{G}^{1} ,\mathbf{G}^{2},\dots ,\mathbf{G}^{B} )$ as the sequence of interaction matrices for different behaviors, with $\mathbf{G}^{b}$ being the interaction matrix for the $b$-th behavior and $\mathbf{G}^{B}$ representing the target behavior. For each behavior, the interaction matrix is binary, indicating whether a user 
$u$ has interacted with an item $i$ under behavior $b$ (1 for interaction, 0 otherwise), formalized as:
\begin{equation}
\label{equation:G_ui_b}
\mathbf{G}_{ui}^{b}= \begin{cases}
  & 1, \text{ If user $u$ has interacted with item $i$ under behavior b ;}  \\
  & 0, \text{ otherwise.} 
\end{cases}
\end{equation}

In this study, the task of multi-behavior recommendation is defined as follows:\\
\begin{itemize}
\item \textbf{Input}: The set of users $\mathcal{U}$ and the set of items $\mathcal{I}$, and user-item interaction matrices $\mathbf{G}=(\mathbf{G}^{1} ,\mathbf{G}^{2},\dots ,\mathbf{G}^{B} )$ for $B$ behaviors.\\
\item \textbf{Output}: The system predicts a similarity score, which indicates the likelihood that a user $u$ will be interacting with an item $i$ under the target behavior $B$.  Items are then recommended to the user based on a descending order of these similarity scores.
\end{itemize}

\begin{figure*}[t]
	\centering
	\includegraphics[width=1.0\linewidth]{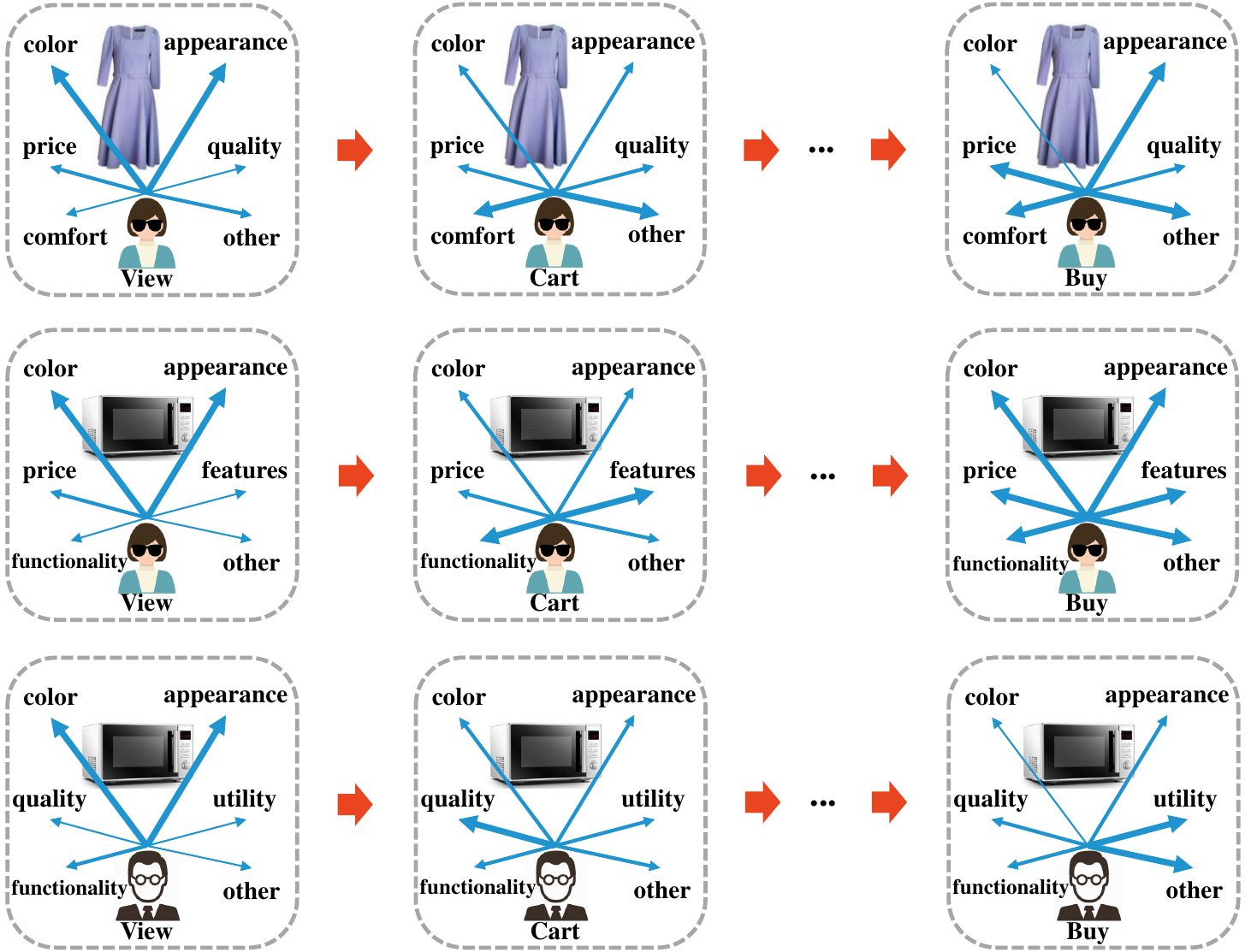}
	\caption{Illustration of user preference for different types of behaviors, where the boldness of each arrow indicates the degree of preference.}
	\vspace{8pt}
	\label{fig:intro}
\end{figure*}

\subsection{Disen-CGCN Model}

In this section, we delve into the details of our Disen-CGCN model. Prior research has shown the benefits of modeling multi-behavior data through cascading behavior relationships, where embeddings from one behavior serve as inputs for the subsequent one, capturing behavioral dependencies in the embedding learning process~\cite{yan2023cascading,cheng2023multi}. 
In real-world scenarios, the factors influencing user decisions vary across behaviors, while the personalized feature transformation between user behaviors also varies from person to person. To illustrate these ideas, we will provide two vivid examples using a shopping platform as the context (see Figure~\ref{fig:intro}). Typically, user behavior in the initial phase is driven by superficial factors such as vision or cost, and as the behavior progresses, users tend to shift their focus towards more specific attributes such as comfort, quality, and functionality. For instance, let's consider a scenario where a woman is shopping for a new "dress" and a "microwave oven". When \textit{browsing} for a dress, her attention may initially be focused on appearance and color, but during the \textit{add-to-cart} phase, preferences shift to other attributes like comfort, and the final \textit{purchase} decision is influenced by the combined effects of different attributes, like appearance, comfort and price, etc. When \textit{browsing} for a microwave oven, users may initially focus on appearance and color due to limited knowledge of the product. As more is learned, users may shift their attention to functionality and features during the \textit{add-to-cart} stage, and ultimately focus on value for money (all factors considered) in their \textit{purchase} decision. Furthermore, we observe another example in the figure where a man is also shopping for a microwave oven. In his case, he initially focuses on color and appearance during \textit{browsing}, but during the \textit{add-to-cart} phase, his attention shifts to quality. Ultimately, his decision to \textit{purchase} the item is driven by other attributes such as utility. This further emphasizes that even when two users are shopping for the same item at the same time, their attention and preferences can differ due to personalized factors and shifting preferences between behavioral transformations. It is evident that each user has their own intrinsic factors that drive their behavior on the platform, resulting in distinct preferences and personalized feature transformations between behaviors.

\begin{figure*}[t]
	\centering
	\includegraphics[width=1.0\linewidth]{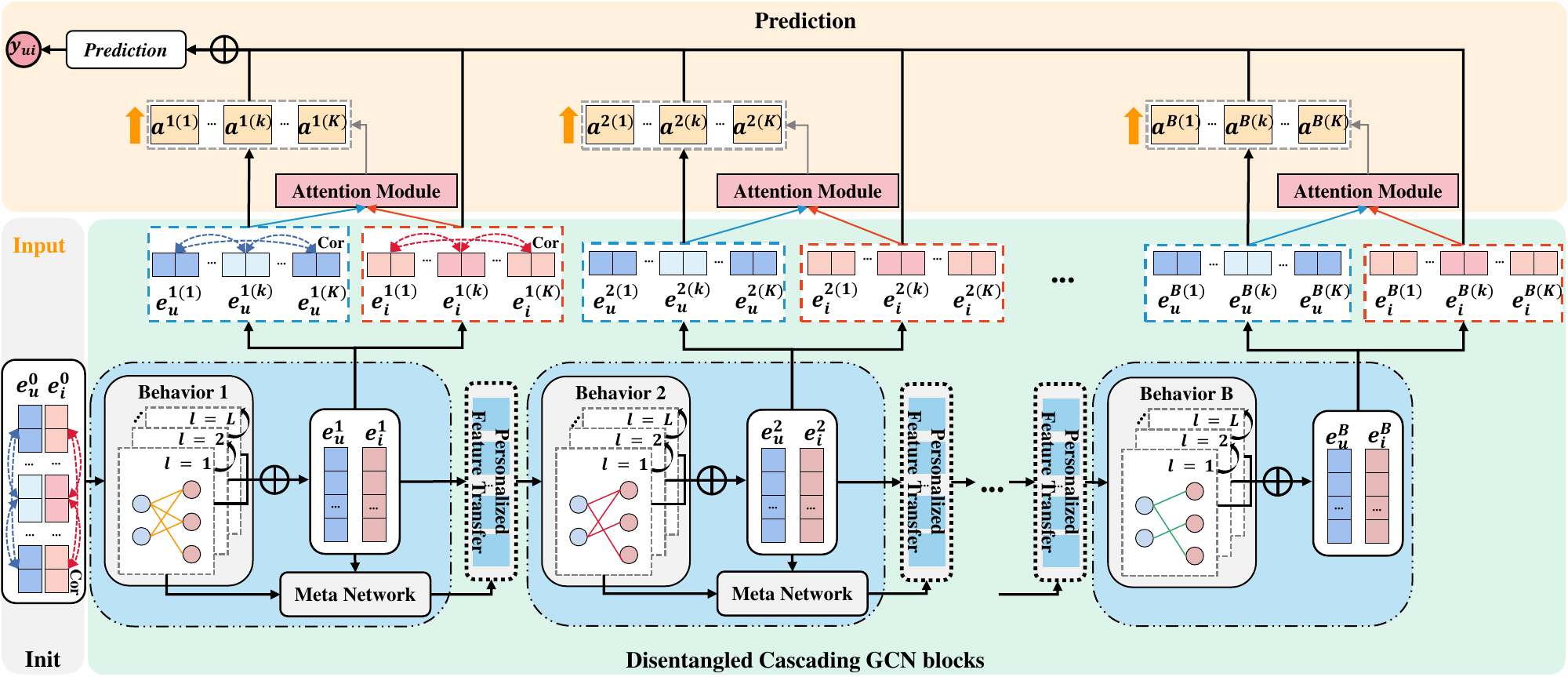}
	\caption{Overview of our Disen-CGCN model.}
	\vspace{8pt}
	\label{fig:Disen-CGCN}
\end{figure*}
The above example highlights two crucial observations. \textit{Firstly}, users exhibit varying preferences for different factors across behaviors, necessitating a method to effectively disentangle and independently evaluate these varying factors for both users and items, as well as to compute users' preferences for different item factors. \textit{Secondly}, in modeling cascading behavior relationships, where one behavior’s embeddings are inputs for the next, the challenge lies in executing personalized feature transformations across behaviors. These considerations form the core problems addressed by our Disen-CGCN model. To tackle the identified challenges, our approach integrates multiple types of interaction data and behavioral cascading relationships in a multi-behavior recommendation model. This model offers a more nuanced understanding of user preferences for various item factors in each behavior and facilitates personalized feature transformations between behaviors. The architecture of our Disen-CGCN, depicted in Figure~\ref{fig:Disen-CGCN}, comprises three key components:
\begin{itemize}
\item \textbf{Embedding Initialization:} This stage involves initializing user and item embeddings, setting the foundation for the entire model's learning process. 
\item \textbf{Disentangled Cascade GCN Blocks:} At the core of our model, we employ the efficient LightGCN model to learn the embeddings of users and items for each behavior. We then apply a disentangled representation technique to disentangle the different factors of user and item representations and ensure that the different factors are independent of each other. Additionally, meta-networks are designed to facilitate the personalized transformation of features for users and items across various behaviors.
\item \textbf{Prediction:} In this final phase, we design an attention mechanism to explicitly model the user's varying preferences for different factors of items across each behavior. This mechanism is crucial in aggregating the influence of all these factors to make the final prediction. To achieve this, we linearly aggregate the embeddings from each behavior, weighted by the attention scores, to predict the user's overall preference for the item. In this way, we can comprehensively capture users' preferences and generate accurate recommendations.
\end{itemize}

%To address these two issues, we leverage multiple types of interaction data and behavioral cascading relationships to design a multi-behavior recommendation model with more fine-grained modeling, which captures the user's preference for different factors of the item in each behavior as well as the personalized feature transformations of the user and the item between the behaviors. The overall structure of our Disen-CGCN is shown in Figure~\ref{fig:Disen-CGCN}, which consists of three important components: 1) \textbf{Embedding Initialization}, this is the learning of the entire model to initialize the embeddings of users and items. 2) \textbf{Disentangled Cascade GCN Blocks}, to obtain the embeddings of users and items in each behavior we utilize the simple and efficient LightGCN model for learning, and subsequently we employ disentangled representation technique and attention mechanism to show the modeled user's preference for different factors of the item. Meanwhile, we also design meta-networks to transform personalized features for users and items between different behaviors.3) \textbf{Embedding Aggregation}, here the embeddings learned from each behavior are aggregated to predict the overall preference of users for items.
\subsubsection{\textbf{Embedding Initialization.}}
Following the existing deep learning-based recommendation methods~\cite{he2017neural,cheng2018aspect,he2020lightgcn,liu2020attribute}, we represent a user $u\in \mathcal{U}$ and an item $i\in \mathcal{I}$ with ID embeddings. They are initialized as $\bm{e_{u}^{0}} \in \mathbb{R}^{d} $ and $\bm{e_{i}^{0}} \in \mathbb{R}^{d} $, where $d$ denotes the embedding size. Let $\bm{P}\in \mathbb{R}^{M\times d}$ and $\bm{Q}\in \mathbb{R}^{N\times d}$ denote the embedding matrices of users and items, with $M$ and $N$ being the total number of users and items, respectively. Each user and item is uniquely represented by a one-hot vector. We use $\bm{ID^{U}} $ and $\bm{ID^{I}} $ to denote the one-hot vector embeddings of all users and items. Formally, the embedding of a user $\bm{u_{m}}$ and an item $\bm{i_{n}}$ is initialized as:
\begin{equation}
\label{equation:init}
\begin{aligned}
\bm{e_{u_{m}}^{0} =P\cdot ID_{m}^{U},} 
\bm{e_{i_{n}}^{0} =Q\cdot ID_{n}^{I},}
\end{aligned}
\end{equation}
where $\bm{ID_{m}^{U}}$ and $\bm{ID_{n}^{I}} $ are the one-hot vectors of users $\bm{u_{m}}$ and items $\bm{i_{n}}$, respectively. In our model,the initial embeddings of users and items serve as the input for the LightGCN model of the first behavior\footnote{It's important to highlight that due to the cascading nature of behaviors in our model, where embeddings learned in one behavior serve as input features for the subsequent behavior, an effective pre-training operation is critical. Inadequately learned embeddings in earlier behaviors can adversely affect the learning process in later behaviors.}. 

\subsubsection{\textbf{Disentangled Cascading GCN blocks.}}
%The goal of the disentangled cascade GCN block is to capture more fine-grained intentional preferences of users for items and personalized feature transitions between users and items, specifically, users have different preferences for items in different behaviors as well as personalized information transfer between users and items between different behaviors. We make full use of the cascade relationship between behaviors, which has been proven to be an effective solution to model multi-behavior recommendation. We first use GCN-based to learn the embeddings of users and items in each behavior, then use disentangled representation techniques and attention mechanisms to capture users' fine-grained intent preferences for items, and finally use meta-networks to perform personalized feature transformations of users and items based on user and item personalization information. Next, we will introduce the disentangled cascade GCN block in four functions.

The module is specifically designed to intricately capture users' detailed preferences for items and facilitate personalized feature transitions between users and items across various behaviors. The GCN blocks harness the cascade relationship among behaviors, a strategy that has proven effective in multi-behavior recommendation modeling~\cite{yan2023cascading,cheng2023multi,meng2023parallel}. Initially, we employ a GCN-based model to learn user and item embeddings for each distinct behavior. Following this, disentangled representation techniques are applied to precisely discern users' nuanced intent preferences for different factors of items. Subsequently, we utilize meta-networks to conduct personalized feature transformations for both users and items, informed by their unique interaction data. This multifaceted approach ensures a comprehensive understanding of user-item interactions across multiple behaviors. In the subsequent sections, we will delve into the four key functions of the disentangled cascade GCN block, elucidating how each contributes to the overall effectiveness of our recommendation model.

\textbf{Single behavior modeling.} To well capture user preferences from each behavior, we adopt the GCN-based recommendation model as backbone to learn user and item embedding based on the user-item bipartite graph built upon the interactions of each behavior, due to great success achieved by GCN-based models in recommendation. A core principle of GCN-based methods is aggregating information from higher-order connections. This involves a recursive process of fusing the embeddings of neighboring nodes and updating the embeddings of the nodes themselves. Since the main goal of this study is to validate the importance of modeling users' finer-grained perferences at factor-level across multi-behaivors, we also adopt LightGCN as the backbone model as in MB-CGCN~\cite{cheng2023multi}. LightGCN simplifies the traditional GCN architecture by removing the transformation matrix and nonlinear activation function, enhancing its efficiency and effectiveness in recommendation tasks. While LightGCN is our primary model, it's important to note that other GCN models such as GTN~\cite{fan2022graph} and Ultra-GCN\cite{mao2021ultragcn} are also suitable for modeling single behaviors.

For the initial embeddings of a user and an item in the first behavior, denoted as  $\bm{e_{u}^{b}} $ and $\bm{e_{i}^{b}}$, LightGCN executes graph convolution operations on the user-item interaction graph of the $b$-th behavior as:
\begin{equation}
\label{equation:GCN}
\begin{aligned}
\bm{e_{u}^{(b,l+1)}} =\sum_{i\in \mathcal{N}_{u}}\frac{1}{\sqrt{\left |\mathcal{N}_{u}  \right | }\sqrt{\left |\mathcal{N}_{i}  \right | } }\bm{e_{i}^{(b,l)}} , \\
\bm{e_{i}^{(b,l+1)}} =\sum_{u\in \mathcal{N}_{i}}\frac{1}{\sqrt{\left |\mathcal{N}_{i}  \right | }\sqrt{\left |\mathcal{N}_{u}  \right | } }\bm{e_{u}^{(b,l)}} ,
\end{aligned}
\end{equation}
where $\bm{e_{u}^{(b,l)}}$ and $\bm{e_{i}^{(b,l)}}$ respectively denote the updated embeddings of user $u$ and item $i$, derived from the $l$ layer under behavior $b$. $\frac{1}{\sqrt{\left |\mathcal{N}_{u}  \right | }\sqrt{\left |\mathcal{N}_{i}  \right | } }$ is the  normalization term, where $\mathcal{N}_{u}$ refers to the set of items that user $u$ has interacted with, and $N_{i}$ denotes the set of users who have interacted with item $i$. After  propagating through $L$ layers, LightGCN integrates the embeddings from each layer to construct the final comprehensive embeddings for users and items. Therefore, the representation of user $u$ and item $i$ learned by LightGCN under behavior $b$ is given by:
\begin{equation}
\label{equation:agg}
\begin{aligned}
\bm{e_{u}^{(b)}}=\sum_{l=0}^{L}\alpha _{l}\bm{e_{u}^{(b,l)}},~~~~
\bm{e_{i}^{(b)}}=\sum_{l=0}^{L}\alpha _{l}\bm{e_{i}^{(b,l)}},
\end{aligned}
\end{equation}
where $\alpha _{l} \ge 0$ is a hyper-parameter controlling the weight of the $l$-th level embedding in the aggregated representation. Typically, for simplicity, we set $\alpha _{l}$  uniformly to $\frac{1}{L+1} $. The embeddings derived from under behavior $b$-th are further refined using disentangled representation techniques to discern and model the impact of different factors on varying user preferences.

\textbf{Disentangled representation learning.} %Intuitively, different types of interaction behaviors reveal different user preferences. \cite{wan2018item} points out that typically the latter behavior in a behavior chain will contain richer signals than the current behavior. However, recent work on modeling multi-behavior data using behavior chains has rarely explored the impact of different factors on different user preferences within each behavior. They usually put a whole embedding in each behavior to represent the user and the item, so that the user's various preferences are entangled and it is not possible to capture for what reasons the user actually interacted with the item in this behavior. Therefore, our goal is to decouple the overall embedding of the user and the item into different factors (e.g., semantic information such as price, color, etc.) in different behaviors and make each factor independent of each other, so that the user's different preferences can be modeled based on the different factors. We use disentangled representation techniques to achieve this goal. In the model, we uniformly divide the user and item embeddings learned in each behavior into K blocks, which is denoted as:
Intuitively, different types of interaction behaviors reflect distinct aspects of user preferences. As~\cite{wan2018item} suggests, later behaviors in a behavior chain typically contain richer signals than earlier ones. Yet, recent approaches to modeling multi-behavior data using behavior chains have not thoroughly examined how different factors influence user preferences within each specific behavior. Common practice involves using a single embedding to represent both the user and the item in each behavior. This approach, however, leads to the entanglement of the user's diverse preferences, making it challenging to discern the specific reasons behind a user's interaction with an item in a particular behavior. To address this, our goal is to disentangle the comprehensive embeddings of both the user and the item into distinct factors (such as semantic attributes like price and color) across various behaviors, ensuring each factor remains independent.  This separation allows for more accurate modeling of user preferences based on these individual factors. 

To accomplish this, we utilize disentangled representation techniques in our model. Specifically, the embeddings of users and items learned in each behavior are uniformly divided into $K$ distinct blocks, represented as:
\begin{equation}
\label{equation:split}
\begin{aligned}
\bm{e_{u}^{(b)}}=[\bm{e_{u}^{(b,1)}},\dots ,\bm{e_{u}^{(b,k)}},\dots,\bm{e_{u}^{(b,K)}}] ,\\
\bm{e_{i}^{(b)}}=[\bm{e_{i}^{(b,1)}},\dots ,\bm{e_{i}^{(b,k)}},\dots,\bm{e_{i}^{(b,K)}}] ,
\end{aligned}
\end{equation}
where $\bm{e_{u}^{(b,k)}}\in \mathbb{R}^{\frac{d}{K}} $ and $\bm{e_{i}^{(b,k)}}\in \mathbb{R}^{\frac{d}{K}}$ denote the $k$-th block of the user and item embeddings, respectively. Each block is assumed to represent a distinct factor, with these factors being independent of each other. This independence allows us to capture the diverse preferences of users for items across different behaviors more effectively.  

However, even with the division of users and items into $K$ blocks of embedding vectors in each behavior, the challenge remains that different factors are still intertwined, leading to issues of information redundancy. To counter this and ensure the independence of factors within the $K$ blocks, we apply distance correlation, which is described as:
\begin{equation}
\label{equation:Lx}
L_{e}=\sum_{k=1}^{K} \sum_{k^{'}=k+1}^{K}dCor(\bm{e^{k}}, \bm{e^{k^{'} }}) ,
\end{equation}
where $\bm{e^{k}}$ and $\bm{e^{k^{'}}}$ denote the embeddings of $k$-th and $k^{'}$-th blocks of the feature vector $\bm{e}$, respectively. The function $dCor(\cdot )$ is a calculates the distance correlation,  defined as: 
\begin{equation}
\label{equation:dcor}
dCor(\bm{x^{k}},\bm{x^{k^{'}}} )=\frac{dCov(\bm{x^{k}},\bm{x^{k^{'}}})}{\sqrt{dVar(\bm{x^{k}})dVar(\bm{x^{k^{'}}})} } ,
\end{equation}
where $dCov(\cdot )$  denotes the distance covariance between two matrices, and $dVar(\cdot )$  represents the distance variance of each matrix.  Detailed explanation of these concepts is provided in~\cite{szekely2007measuring}.

In our Disen-CGCN model, we divide the embeddings of users and items for all behaviors into $K$ blocks. Theoretically, corresponding blocks in different behaviors should represent identical factors. For example, if the first block in the first behavior's embedding signifies the "color" factor, then the corresponding first block in the embeddings of all subsequent behaviors should also represent the "color" factor.   It is critical to address  the alignment of these factors across corresponding blocks in each behavior. Initially, we explored  a block-wise contrastive learning approach, where blocks in corresponding positions across two behaviors  were paired as positive pairs, and blocks in different positions  were considered negative pairs. The fundamental concept in contrastive learning is to bring positive pairs closer together and distance the negative pairs. However, this approach becomes increasingly complex and computationally demanding with more blocks and behaviors due to the necessity of pairwise alignment.

To overcome this complexity, we leveraged the inherent structure of the behavior chain. By ensuring that the $K$ block embeddings of users and items in the first behavior are kept independent and then individually transferring these embeddings to subsequent behaviors through block-wise feature transformation, we maintain both independence and factor consistency across behaviors. It's important to note that these independent $K$ block embeddings are utilized as the initial embeddings for the GCN-based collaborative filtering learning in the subsequent behavior. Because the whole learning process involving user and item embedding learning operates on an element-wise basis, this ensures that these transformed $K$ block embeddings remain independent. The process of maintaining independent block embeddings is  repeated in each subsequent behavior in the model.

Inspired by this, we focus on constraining the independence of the $K$ block embeddings only in the first behavior, ensuring each pair of embeddings is independent.  Given that our model also integrates personalized feature transformations, which are elaborated on in subsequent paragraphs, it becomes essential to facilitate the feature transformation from the first to the second behavior (as described in Eq.\ref{equation:T}). To aid this process, we also apply distance correlation constraints to the initial $K$ block embeddings of both users and items (namely, $e_{u}^{(0)}$ and $e_{i}^{(0)}$).  The independence constraint in our model is thus formulated as:

\begin{equation}
\label{equation:D_loss}
\ell  _{d} = L_{\bm{e_{u}^{(0)}}} +L_{\bm{e_{i}^{(0)}}} +L_{\bm{e_{u}^{(1)}}} +L_{\bm{e_{i}^{(1)}}}.
\end{equation}

\textbf{Personalized feature transformation.} Our Disen-CGCN model is specifically crafted to uncover more detailed user preferences within multi-behavior data. While above components in our model focus on exploring the varied preferences users have for items in each behavior, it's equally important to consider the preference and relationship transfers  between different behaviors. For instance, a user may focus more on a particular factor in one behavior, suggesting that the feature input from the previous behavior into the current one should retain more information about this specific factor. Thus, the information transferred between behaviors is inherently individualized. Similarly, for items, the factors attracting users' interest vary across behaviors, necessitating that item information transfer between behaviors aligns more closely with the users' personalized preferences in each specific behavior. To address this need for personalized feature transformation between different behaviors, we drew inspiration from works such as ~\cite{xia2021graph,zhu2022personalized,chen2023heterogeneous}, and designed meta-networks that facilitate customized feature extraction between behaviors. This approach enables our model to better capture the varying preferences of users, offering a more personalized and accurate representation of their interests across various behaviors.

To facilitate the transition of personalized feature mappings for users and items from the current behavior $(b)$ to the next behavior $(b+1)$ in our Disen-CGCN model, we begin by extracting meta-knowledge from the user and item features present in behavior $b$. This step is crucial for retaining the most personalized and salient features of both users and items. Note that during this feature transformation process, we must maintain the independence of the $K$ block embeddings from behavior $b$. To achieve this, we handle each of the $K$ block embeddings individually. The meta-knowledge in the $b$-the block is defined as follows:

\begin{equation}
\label{equation:T}
\begin{aligned}
\bm{T_{u}^{(b,k)}}=\bm{e_{u}^{(b,k)}}||\frac{\sum_{i\in \mathcal{N}_{u}}\bm{e_{i}^{(b-1,k)}}}{\sqrt{\left |\mathcal{N}_{u}  \right | \left |\mathcal{N}_{i}  \right | } }  ,\\
\bm{T_{i}^{(b)(k)}}=\bm{e_{i}^{(b,k)}}||\frac{\sum_{u\in \mathcal{N}_{i}}\bm{e_{u}^{(b-1,k)}}}{\sqrt{\left |\mathcal{N}_{i}  \right | \left |\mathcal{N}_{u}  \right | } }  , 
\end{aligned}
\end{equation}
where $\bm{T_{u}^{(b,k)}}\in \mathbb{R}^\frac{2d}{K} $ and $\bm{T_{i}^{(b,k)}}\in \mathbb{R}^\frac{2d}{K} $ denote the meta-knowledge of the $k$-th block embedding of the user and item in behavior $b$, respectively. Meanwhile, $\bm{e_{u}^{(b,k)}}\in \mathbb{R}^\frac{d}{K} $ and $\bm{e_{i}^{(b,k)}}\in \mathbb{R}^\frac{d}{K} $ represent the $k$-th block embedding of the user and item learned in behavior $b$. Additionally,  $\sum_{i\in \mathcal{N}_{u} }\bm{e_{i}^{(b-1,k)}}\in  \mathbb{R}^{\frac{d}{K} }$ and $\sum_{u\in \mathcal{N}_{i} }\bm{e_{u}^{(b-1,k)}}\in  \mathbb{R}^{\frac{d}{K} }$ correspond to the aggregated embeddings of the first-order neighbor around the user and item learned from the previous behavior, respectively. The term $\frac{1}{\sqrt{\left |\mathcal{N}{u} \right | \left |\mathcal{N}{i} \right | }}$ is used as the  normalization factor. The notation $||$ denotes the concatenation operation.

Our model distinctively employs  the $k$-th block embeddings of both the user's and item's own nodes, as well as the $k$-th block embeddings of their first-order neighbors during graph convolution, to serve as meta-knowledge. This strategy, combining features learned from interactions in the current behavior with those directly inherited from the previous behavior (represented by the embedding aggregation of the first-order neighbors), is designed to more accurately capture the personalized features of users and items.  Notably, while studies like \cite{xia2021graph,chen2023heterogeneous} typically utilize the embeddings of the node itself and its neighbors after graph convolution  (essentially, nodes' features and those of their neighbors learned in the current behavior), our findings suggest that this method is less effective for our model, as demonstrated and discussed in the experimental section of our study (refer to Section~\ref{section:mks}). The improved performance observed in our approach might be attributed to the feature transformation that considers both current and previous behaviors, functioning like the residual connections in CRGCN~\cite{yan2023cascading}.

%We use the $k$-th block embeddings of the user's and item's own nodes and the $k$-th block embeddings of the first-order neighbors during graph convolution as meta-knowledge, which can represent the most personalized features of the user and item. Note that \cite{xia2021graph,chen2023heterogeneous} typically utilize the self node as well as its neighbors after graph convolution is complete, which was found to be less than optimal in our model, and we will show the results of both approaches in the experimental section.

In the next, we leverage the gathered meta-knowledge in our meta-network to generate  user and item personalization transformation matrices. The meta-network is defined as:
\begin{equation}
\label{equation:meta_network}
\begin{aligned}
\bm{M_{u}^{(b,k)}}=f(\bm{T_{u}^{(b,k)}} ) ,\\
\bm{M_{i}^{(b,k)}}=f(\bm{T_{i}^{(b,k)}} ) ,
\end{aligned}
\end{equation}
where $f(\cdot )$ represents the meta-network, specifically a two-layer feed-forward neural network. This network is designed to process meta-knowledge as its input. Its output consists of two distinct types of matrices: 
$\bm{M_{u}^{(b,k)}}\in \mathbb{R}^{\frac{d}{K}\times  \frac{d}{K}} $ and $\bm{M_{i}^{(b,k)}}\in \mathbb{R}^{\frac{d}{K}\times  \frac{d}{K}} $. These matrices are personalized transformation matrices and are embedded in the $k$-th block of the respective user and item for a given behavior $b$. The primary function of the meta-network is to generate these customized transformation matrices,  specifically designed to align with the distinct, personalized characteristics of each user and item.  This tailored process is essential for achieving effective personalized feature transformation. The operational methodology of this process is encapsulated in the subsequent formula:
\begin{equation}
\label{equation:PTF}
\begin{aligned}
\bm{e_{u}^{(b+1,0,k)}}=\bm{M_{u}^{(b,k)}} \bm{e_{u}^{(b,k)}}  ,\\
\bm{e_{i}^{(b+1,0,k)}}=\bm{M_{i}^{(b,k)}} \bm{e_{i}^{(b,k)}}  ,
\end{aligned}
\end{equation}
where $\bm{e_{u}^{(b+1,0,k)}}\in \mathbb{R}^{\frac{d}{K} } $ and $\bm{e_{i}^{(b+1,0,k)}}\in \mathbb{R}^{\frac{d}{K} } $ denote the initial embeddings of the $k$-th block of the user and item in the $(b+1)$-th behavior, respectively\footnote{Note that ``0" in $\bm{e_{u}^{(b+1,0,k)}}$ and $\bm{e_{i}^{(b+1,0,k)}}$ represents the initial embedding of users and items in the LightGCN model for the $k$-th block in the $b+1$ behavior. For the $l$-th layer, the representation should be $\bm{e_{i}^{(b+1,l,k)}}$ or $\bm{e_{i}^{(b+1,l,k)}}$.}. With this module, the Disen-CGCN model not only captures the nuanced user preferences within each distinct behavior but also effectively facilitates personalized feature transformation for users and items across varying behaviors.

To enable personalized feature transformation and address the challenge of preference transfer between users and items across different behaviors, we introduce a meta-network that extracts the meta-knowledge, which represents the personalized information of users and items in each behavior. This meta-network generates a weight matrix that captures the personalized feature information for users and items. When transitioning from the current behavior to the next behavior, the personalized feature transformation network utilizes the weight matrix obtained from the meta-network. This weight matrix holds the personalized feature information of users and items. By applying this weight matrix, we transform the personalized preference information of users and items, respectively, to align with the next behavior. The transformed results serve as the initial embeddings for the GCN encoder in the next behavior. 

For example, let's consider two users, $u_1$ and $u_2$, who have different intrinsic preferences regarding the price factor (post-disentangling factor). User $u_1$ prefers items with low prices, while user $u_2$ prefers items with high prices. Initially, the GCN encoders model the preference information of these users within their current behavior. The meta-network extracts the meta-knowledge of these two users and generates a weight matrix for the personalized feature transformation network. Importantly, the intrinsic preferences of users (low or high price) are relatively fixed and remain unchanged during the behavioral transformation process. This means that the personalized information of each user retains their intrinsic preference information throughout the transformation. For instance, user $u_1$'s personalized information still reflects their preference for items with low prices, while user $u_2$'s personalized information continues to indicate their preference for items with high prices. These transferred results are then utilized to further model subsequent behaviors using the GCN encoder.

\subsubsection{\textbf{Prediction}.}
% \textbf{Intentional preference modeling}: 
Users typically exhibit varying preferences across different behaviors. For instance, in the \textit{browsing} phase, a user might focus more on an item's appearance, whereas in later stages like \textit{adding to cart} or \textit{purchasing}, the price may become more significant. To capture these behavior-specific preferences, the Disen-CGCN model employs an attention mechanism. Specifically, for item $i$ in behavior $b$, the user 
$u$'s preference for the $k$-th factor is calculated using the following equation:
\begin{equation}
\label{equation:h}
{\hat{a} _{u}^{(b,k)}}=\bm{h_{u}^{(b)^{T} }} Tanh(\bm{W_{u}^{(b)}}[\bm{e_{u}^{(b,k)}};\bm{e_{i}^{(b,k)}}]+\bm{b_{u}^{(b) }} ) ,
\end{equation}
where $\bm{e_{u}^{(b,k)}}\in  \mathbb{R}^{\frac{d}{K} }$ and $\bm{e_{i}^{(b,k)}}\in  \mathbb{R}^{\frac{d}{K} }$ denote the $k$-th embeddings of the user and the item under behavior $b$, respectively. $\bm{W_{u}^{(b)}}\in  \mathbb{R}^{\frac{2d}{K}\times d } $ and $\bm{b_{u}^{(b)}}\in  \mathbb{R}^{d } $ denote the weight matrix and bias vector, mapping  these embeddings to the hidden layer for behavior  $b$. $\bm{h_{u}^{(b)^{T}}}\in  \mathbb{R}^{d} $ maps the hidden layer to the output attention weight under behavior $b$. $[\bm{e_{u}^{(b,k)}};\bm{e_{i}^{(b,k)}}]$ represents the concatenation of $\bm{e_{u}^{(b,k)}}$ and $\bm{e_{i}^{(b,k)}}$, with $tanh$ as the activation function.

Following neural attention network principles, $\bm{\hat{a} _{u}^{(b,k)}}$ is normalized using a $softmax$ function, transforming the attention weights into probability distributions.  However, since we are modeling user preferences across different behaviors, the attention scores are applied only to the user's $K$ block embeddings.  This can result in significant variations in the magnitude of user and item embeddings across behaviors. Inspired by \cite{liu2019user},  we adjust the normalized weights with a factor $\rho$. In our model, the final attention score is computed as:
\begin{equation}
\label{equation:attention}
{a_{u}^{(b,k)}}=\rho  \cdot \frac{exp(\bm{\hat{a} _{u}^{(b,k)}})}{\sum_{k^{'}=1 }^{K} exp(\bm{\hat{a} _{u}^{(b,k^{'})}})} .
\end{equation}

Unlike \cite{liu2019user}, where $\rho$ is set as the dimension of the weight vector, we treat $\rho$  as a hyper-parameter to enhance its influence on the model. In our model, attention weights play a crucial role in measuring user preferences for various factors across different behaviors. To achieve the optimal recommendation performance, the model employs a linear aggregation of learned embeddings, effectively incorporating these attention weights. This approach is designed to reflect the user's diverse preferences across all behaviors.

Following the method described in \cite{cheng2023multi}, we utilize a straightforward yet effective method to first aggregate the embeddings. Then, it predicts a user $u$'s preference for an item $i$ using the following formula:
\begin{equation}
\label{equation:predict}
\hat{y} _{ui}  =\sum_{k=1}^K (\sum_{b=1}^B {a_{u}^{(b,k)}}\bm{e_{u}^{(b,k)}})^T (\sum_{b=1}^B \bm{e_{i}^{(b,k)}}) . 
\end{equation}

In this equation, the term $\sum_{b=1}^B {a_{u}^{(b,k)}}\bm{e_{u}^{(b,k)}}$ represents the aggregation of user $u$'s preferences for the $k$-th factor, taking into account the varying degrees of preference across different behaviors.  Accordingly, the product $(\sum_{b=1}^B {a_{u}^{(b,k)}}\bm{e_{u}^{(b,k)}})^T  (\sum_{b=1}^B \bm{e_{i}^{(b,k)}})$  estimates the preference score for the 
$k$-th factor of the target item by leveraging information from all behaviors.  The final prediction is derived by combining these calculated scores for all $K$ factors. The recommender system will recommend items to the user in the order of their scores from highest to lowest.

\subsection{Model Training}
Our objective is to optimize the Disen-CGCN model for top-$n$ recommendation. This involves recommending a set of $n$ items to a user, ranked by their predicted preference scores. To achieve this, we adopt the pairwise Bayesian personalized ranking  loss (BPR) function \cite{rendle2012bpr}. In this framework, each training sample consists of a user $u$, a positive item $i^{+}$ with which the user has interacted, and a negative item $i^{-}$with which the user has not interacted. The BPR loss function operates under the assumption that the user $u$'s score for the positive item $i^{+}$ should be higher than for the negative item  $i^{-}$. The objective function is formulated as follows:
\begin{equation} 
\label{equation:BPR}
\ell _{rec}=\sum_{(u,i^{+},i^{-})\in O  }-ln\sigma(\hat{y} _{ui^{+}}-\hat{y} _{ui^{-}} )+\lambda \left \| \Theta  \right \| _{2}^{2},
\end{equation}
here, $ O =\{(u,i^{+},i^{-})|(u,i^{+})\in R^{+},(u,i^{-})\in R^{-} \} $ represents the training dataset, where $R^{+}$ and $R^{-}$ are sets of items that user $u$ has interacted with and not interacted with, respectively.  The function $\sigma (\cdot )$ is the sigmoid function. $\Theta$ encompasses all trainable parameters within the model.  We apply $L_{2}$ regularization, controlled by the coefficient 
$\lambda$ , to prevent overfitting. The total loss of the Disen-CGCN model is computed as::
\begin{equation} 
\label{equation:Loss}
\ell =\ell _{rec}+\beta\ell _{d} ,
\end{equation}
where $\beta$ is a hyper-parameter that determines the weight of the disentangled representation module within the overall loss function.
\section{EXPERIMENT} \label{experiment}
In this section, we conduct extensive experiments on two real-world datasets to evaluate the performance of Disen-CGCN.  Additionally, we provide visual analyses to explore user preferences across different behaviors and assess the impact of key components and hyper-parameters on the model's effectiveness.  Our empirical study mainly answers the following research questions:

\begin{itemize}
    \item \textbf{RQ1}: How does Disen-CGCN perform against benchmark methods?
    
    \item \textbf{RQ2}: Can Disen-CGCN effectively capture and represent the variance in user preferences for items across different behaviors?
    
    \item \textbf{RQ3}: What is the contribution of each key component in enhancing Disen-CGCN’s performance?

    \item \textbf{RQ4}: How do different hyper-parameters influence the overall performance of Disen-CGCN?
\end{itemize}

\subsection{Experiment Settings}

% \usepackage{booktabs}
\begin{table}[]
\centering
\caption{Basic statistics of the experimental datasets.}
\label{data_describe}
\begin{tabular}{cccccc}
\toprule
Dataset & \#User  & \#Item  & \#Buy    & \#Cart   & \#View     \\ \midrule
Beibei  & 21,716 & 7,997  & 304,576 & 642,622 & 2,412,586 \\
Tmall   & 15,449 & 11,953 & 104,329 & 195,476 & 873,954   \\ \bottomrule
\end{tabular}
\end{table}

\subsubsection{\textbf{Dataset.}}
Our experiments were conducted on two publicly accessible datasets: Beibei\footnote{https://www.beibei.com/} and Tmall\footnote{https://www.tmall.com/}. We describe the two datasets as follows:

\begin{itemize}
    \item \textbf{Beibei}:This dataset is sourced from Beibei, China's largest retail e-commerce platform specializing in baby products.  Covering the period from June 1, 2017, to June 30, 2017, it records the interactions of 21,716 users with 7,977 items. The dataset categorizes user behavior into three types: View, Cart, and Buy. It's important to note that the Beibei platform requires users to follow a strict sequential pattern for making a purchase: they first view products, then add the ones they are interested in to their cart, and finally proceed to purchase these items (i.e., view -> cart -> buy).

    \item \textbf{Tmall}: This dataset was collected from Tmall, one of China's largest e-commerce platforms, this dataset comprises interactions between 15,449 users and 11,953 products. It follows the same categorization of user behaviors as the Beibei dataset, including \textit{View}, \textit{Cart}, and \textit{Buy}.
   
\end{itemize}

For both datasets, we adhere to the methodology of previous studies by processing duplicate interactions and retaining only the earliest occurrence of each interaction \cite{gao2019learning,jin2020multi}. The detailed information about these two datasets, as used in our experiments, is presented in Table~\ref{data_describe}.

\subsubsection{\textbf{Evaluation Protocols.}}
We employ the leave-one-out strategy in evaluation, which has been widely used in previous studies~\cite{gao2019learning,jin2020multi,xia2021graph}. Specifically, for each user, we assign the  latest interactions and all the uninteracted items as the test set, while using all other  interactions for training. In the evaluation phase, all items in the test set are ranked based on the scores predicted by the recommendation model, and the top-$n$ items are selected to assess the model's performance.  In order to evaluate the performance of the top-$n$ recommendation, we utilize two standard metrics  Recall and NDCG, which are common in evaluating recommendation systems.

\subsubsection{\textbf{Baselines.}}
We compare our Disen-CGCN model with several competitive recommendation models, including representative single-behavior models and recently advanced multi-behavior models. We describe these models below.

Single-behavior models:
\begin{itemize}
    \item \textbf{MF-BPR~\cite{rendle2012bpr}}: This method is a matrix decomposition approach that uses Bayesian Personalized Ranking Loss Optimization for the top-n recommendation task. It is widely adopted as a baseline for new models. It is a single-behavior based recommendation model,  utilizing only the user-item interaction data from the target behavior.

    \item \textbf{NeuMF\cite{he2017neural}}: This method combines collaborative filtering and neural networks to capture complex interactions between users and items. It uses a shallow generalized decomposition model and a deep multilayer perceptron.

    \item \textbf{LightGCN~\cite{he2020lightgcn}}: This is a GCN-based recommendation model that leverages higher-order connections in the user-item bipartite graph. It simplifies the architecture by retaining only the core neighbor aggregation part and removes the transformation matrix and nonlinear activation function, resulting in a simple yet effective model that significantly improves performance.

    \item \textbf{DGCF~\cite{wang2020disentangled}}: This approach models the intent distribution of each user-item interaction and iteratively improves the intent-aware interaction graph and representation. The technique of disentangled representation is used to encourage the independence of different intentions.
   
\end{itemize}

Multi-behavior models:
\begin{itemize}
    \item \textbf{RGCN~\cite{schlichtkrull2018modeling}}: This method utilizes different edge types in the graph to distinguish relationships between nodes, designing a separate propagation layer for each edge type. As this method is applicable to graphs with isomorphic edges of nodes, it can model multi-behavior recommendations using multi-behavior data.

    \item \textbf{GNMR~\cite{xia2021multi}}: This method explores dependencies between different behaviors after recursive embedding propagation on a unified multi-behavior interaction graph. It introduces a novel relationship aggregation network to model interaction heterogeneity.

    \item \textbf{NMTR~\cite{gao2019learning}}: This method employs a cascading neural network model tailored for multi-behavior recommendations. Utilizing NeuMF as its backbone, it predicts the interaction score for each behavior, subsequently passing this score from one behavior to the next. In addition, it also incorporates multi-task learning to jointly optimize the model parameters.

    \item \textbf{MBGCN~\cite{jin2020multi}}: It is a GCN-based multi-behavior recommendation model, MBGCN learns user preferences on a unified multi-behavior interaction graph, considering the different impacts of multiple behaviors on the target behavior. Additionally, the model enhances item embedding learning through item-item propagation.

     \item \textbf{CRGCN~\cite{yan2023cascading}}: This approach leverages cascading relationships between behaviors and designs a cascading residual network to model multi-behavior data. Features learned from previous behaviors are passed to the current behavior, with residual connections designed during the feature passing process.  The method utilizes multi-task learning for joint optimization.

    \item \textbf{MB-CGCN~\cite{cheng2023multi}}: This is a recently proposed GCN-based multi-behavior recommendation model. MB-CGCN captures behavioral dependencies in behavioral chains for embedding learning. Embeddings learned from one behavior are feature-transformed to serve as input for the next behavioral embedding learning.
   
\end{itemize}

\subsubsection{\textbf{Hyper-parameter Settings.}}
Our Disen-CGCN model is implemented using Tensorflow~\footnote{https://www.tensorflow.org.}.  In all experiments, we set the embedding size to 64 and fix the mini-batch size at 1024. We employ the Adam optimizer for optimization. The model parameters are initialized using the Xavier method. The learning rate is adjusted within the range of $\{1e^{-2}, 1e^{-3}, 1e^{-4}\}$. Both the $L_{2}$ regularization factor $\lambda $ and the distance correlation $\beta$ are explored within the range of $\{1e^{+1}, 1e^{0},\cdots ,1e^{-5}\}$. We search for the optimal number of disentangled factors $K$ and attention enlargement in the set  $\{1,2,4,8\}$.The number of GCN layers per behavior is searched among $\{1,2,3,4\}$.  Following previous work, we utilize an early stopping strategy. For behavioral chains, we perform in the order as suggested in~\cite{yan2023cascading}: view -> cart -> buy. The best results for the three behavioral GCN layers on the Tmall and Beibei datasets are obtained using $\{3,4,2\}$ and $\{3,4,3\}$ layers, respectively.  For other benchmark models, we use their official open-source codes. To ensure a fair comparison, we carefully tune key parameters to achieve the best performance of these models.

\begin{table}[]
\centering
\caption{Overall performance comparisons on Beibei dataset.}
\label{Beibei_Performance}
\resizebox{0.8\textwidth}{!}{%
\begin{tabular}{l|cc|cc|cc}
\hline
\textbf{Method} & \textbf{R@10} & \textbf{N@10} & \textbf{R@20} & \textbf{N@20} & \textbf{R@50} & \textbf{N@50} \\ \hline\hline
MF-BPR & 0.0189 & 0.0047 & 0.0534 & 0.0244 & 0.1015 & 0.0334 \\ 
NeuMF & 0.0230 & 0.0137 & 0.0737 & 0.0292 & 0.1403 & 0.0407 \\ 
LightGCN & 0.0386 & 0.0203 & 0.0628 & 0.0264 & 0.1298 & 0.0395 \\
DGCF & 0.0402 & 0.0198 & 0.0702 & 0.0272 & 0.1413 & 0.0412 \\ \hline
RGCN & 0.0359 & 0.0187 & 0.0687 & 0.0273 & 0.1311 & 0.0368 \\ 
GNMR & 0.0414 & 0.0222 & 0.0732 & 0.0281 & 0.1388 & 0.0377 \\ 
NMTR & 0.0431 & 0.0192 & 0.0781 & 0.0298 & 0.1451 & 0.0391 \\ 
MBGCN & 0.0472 & 0.0263 & 0.0794 & 0.0331 & 0.1495 & 0.0454 \\ 
CRGCN & 0.0523 & 0.0245 & 0.0895 & 0.0352 & 0.1696 & 0.0490 \\ 
MB-CGCN & {\uline{0.0580}} & {\uline{0.0288}} & {\uline{0.0995}} & {\uline{0.0392}} & {\uline{0.1933}} & {\uline{0.0577}} \\ 
\textbf{Disen-CGCN} & \textbf{0.0620*} & \textbf{0.0314*} & \textbf{0.1055*} & \textbf{0.0423*} & \textbf{0.2044*} & \textbf{0.0617*} \\ \hline
Improvement & 6.99\% & 8.89\% & 5.97\% & 7.80\% & 5.74\% & 7.01\% \\ \hline
\end{tabular}%
}
\begin{tablenotes}
    \item The symbol * denotes that the improvement is significant with $p-value < 0.05$ based on a two-tailed paired t-test.
\end{tablenotes}

\end{table}

% % Please add the following required packages to your document preamble:
% % \usepackage{graphicx}
% % \usepackage[normalem]{ulem}
% % \useunder{\uline}{\ul}{}
% \begin{table}[]
% \centering
% \caption{Overall performance comparisons on Beibei dataset.}
% \label{Beibei_Performance}
% \resizebox{0.8\textwidth}{!}{%
% \begin{tabular}{l|cc|cc|cc}
% \hline
% \textbf{Method} & \textbf{R@10} & \textbf{N@10} & \textbf{R@20} & \textbf{N@20} & \textbf{R@50} & \textbf{N@50} \\ \hline\hline
% MF-BPR & 1.89 & 0.47 & 5.34 & 2.44 & 10.15 & 3.34 \\ 
% NeuMF & 2.30 & 1.37 & 7.37 & 2.92 & 14.03 & 4.07 \\ 
% LightGCN & 3.86 & 2.03 & 6.28 & 2.64 & 12.98 & 3.95 \\ 
% DGCF & 4.02 & 1.98 & 7.02 & 2.72 & 14.13 & 4.12 \\ \hline
% RGCN & 3.59 & 1.87 & 6.87 & 2.73 & 13.11 & 3.68 \\ 
% GNMR & 4.14 & 2.22 & 7.32 & 2.81 & 13.88 & 3.77 \\ 
% NMTR & 4.31 & 1.92 & 7.81 & 2.98 & 14.51 & 3.91 \\ 
% MBGCN & 4.72 & 2.63 & 7.94 & 3.31 & 14.95 & 4.54 \\ 
% CRGCN & 5.23 & 2.45 & 8.95 & 3.52 & 16.96 & 4.90 \\ 
% MB-CGCN & {\uline{5.80}} & {\uline{2.88}} & {\uline{9.95}} & {\uline{3.92}} & {\uline{19.33}} & {\uline{5.77}} \\ 
% \textbf{Disen-CGCN} & \textbf{6.20} & \textbf{3.14} & \textbf{10.55} & \textbf{4.23} & \textbf{20.44} & \textbf{6.17} \\ \hline
% Improvement & 6.99\% & 8.89\% & 5.97\% & 7.80\% & 5.74\% & 7.01\% \\ \hline
% \end{tabular}%
% }
% \end{table}

% \usepackage{graphicx}
% \usepackage[normalem]{ulem}
\begin{table}[]
\centering
\caption{Overall performance comparisons on Tmall dataset.}
\label{Tmall_Performance}
\resizebox{0.8\textwidth}{!}{%
\begin{tabular}{l|cc|cc|cc}
\hline
\textbf{Method} & \textbf{R@10} & \textbf{N@10} & \textbf{R@20} & \textbf{N@20} & \textbf{R@50} & \textbf{N@50} \\ \hline\hline
MF-BPR & 0.0081 & 0.0037 & 0.0255 & 0.0153 & 0.0396 & 0.0192 \\
NeuMF & 0.0241 & 0.0131 & 0.0317 & 0.0155 & 0.0495 & 0.0195 \\ 
LightGCN & 0.0410 & 0.0246 & 0.0550 & 0.0281 & 0.0807 & 0.0332 \\ 
DGCF & 0.0404 & 0.0234 & 0.0533 & 0.0263 & 0.0784 & 0.0312 \\ \hline
RGCN & 0.0219 & 0.0111 & 0.0325 & 0.0127 & 0.0414 & 0.0161 \\ 
GNMR & 0.0361 & 0.0218 & 0.0609 & 0.0265 & 0.0969 & 0.0337 \\ 
NMTR & 0.0280 & 0.0144 & 0.0639 & 0.0307 & 0.1041 & 0.0385 \\ 
MBGCN & 0.0511 & 0.0295 & 0.0692 & 0.0352 & 0.1117 & 0.0459 \\ 
CRGCN & 0.0859 & 0.0444 & 0.1374 & 0.0668 & 0.2330 & 0.0857 \\ 
MB-CGCN & {\uline{0.1317}} & {\uline{0.0686}} & {\uline{0.2007}} & {\uline{0.0859}} & {\uline{0.3226}} & {\uline{0.1101}} \\ 
\textbf{Disen-CGCN} & \textbf{0.1469*} & \textbf{0.0761*} & \textbf{0.2172*} & \textbf{0.0938*} & \textbf{0.3430*} & \textbf{0.1186*} \\ \hline
Improvement & 11.50\% & 10.99\% & 8.23\% & 9.18\% & 6.32\% & 7.80\% \\ \hline
\end{tabular}%
}
\begin{tablenotes}
    \item The symbol * denotes that the improvement is significant with $p-value < 0.05$ based on a two-tailed paired t-test.
\end{tablenotes}
\end{table}

\subsection{Overall Performance (RQ1)}
In this section, we present a performance comparison between our Disen-CGCN model and all benchmark models. The results on both the Beibei and Tmall datasets are reported in Table~\ref{Beibei_Performance} and Table~\ref{Tmall_Performance}, respectively. In these tables,   we highlight the best results in bold and the second best in underline. 
Our findings indicate that multi-behavior approaches generally outperform single-behavior approaches across all evaluation metrics, demonstrating the efficacy of leveraging multi-behavior interaction data to capture more nuanced user preferences.  Our Disen-CGCN, specifically, excels in capturing fine-grained user preferences using multi-behavior data, achieving superior performance over all baseline models. For both recommender system evaluation metrics, comparing the performance of the best baseline model among top-$n$ ($n$=\{10, 20, 50\}) items in different ranges, Disen-CGCN improves performance by an average of 7.07\% and 9.00\% in the Beibei and Tmall datasets, respectively.  This underscores the effectiveness of our model.

For recommendation models using a single behavior data, the MF-BPR simply models user-item interactions with an inner product, limiting its ability to capture complex user-item relationships.  NeuMF combines collaborative filtering with neural networks, outperforming MF-BPR by modeling nonlinear relationships through a multilayer perceptron. Compared with NeuMF and MF-BPR, LightGCN further advances performance by exploiting high-order information on the user-item bipartite graph, highlighting the benefits of GCN models. DGCF, which is also a GCN-based model, employs disentangled representation techniques for user intent modeling and achieves the best performance among single-behavior models.

For multi-behavior recommendation models, RGCN, although modeling different types of behaviors, does not effectively differentiate the importance of these behaviors, leading to relatively poorer performance. GNMR and MBGCN, which consider the contribution of different behaviors more effectively, outperform RGCN. In particular, MBGCN excels by modeling item-item relationships and capturing semantic behavior information. NMTR, CRGCN, and MB-CGCN all take into account cascading relationships among multiple behaviors. This approach involves transferring information, such as predicted scores or user/item embeddings, from one behavior to the next. Among them, the NMTR model outperforms GNMR but falls short of MBGCN. This is largely due to MBGCN's effective use of a GCN-based structure and its emphasis on item-item relationships. The NMTR model operates on the principle that a user engaging in multiple behaviors on an item does so within a short timeframe. Consequently, it keeps the embedding vectors for users and items constant across behaviors, with the primary information being the user's predicted score for the item in each behavior. When comparing CRGCN and MB-CGCN, both models pass the embedding information learned in each behavior to the subsequent one, allowing for further refinement of user and item embeddings. This process makes CRGCN and MB-CGCN more nuanced than NMTR in modeling cascaded multi-behavior relationships. MB-CGCN, in particular, advances this approach by replacing the residual links in inter-behavioral transfers with feature transformations. This strategy not only preserves the diversity of information across different behaviors but also filters out potential noise from previous behaviors. Additionally, MB-CGCN aggregates embeddings learned from various behaviors to enhance prediction accuracy, effectively capturing a broader range of behavioral data. As a result, MB-CGCN demonstrates superior performance.

Our Disen-CGCN model employs a GCN-based cascade structure akin to that of MB-CGCN, enabling it to leverage cascade relationships effectively. However, Disen-CGCN distinguishes itself by modeling user preferences with greater precision than MB-CGCN. It achieves this through the use of a disentangled representation technique, which separates various user preference factors across different behaviors. Additionally, an attention mechanism is utilized to estimate the degree of user preference for these factors within the current behavior. Moreover, Disen-CGCN incorporates a meta-network between two GCN blocks, facilitating personalized feature transformation for users and items. This intricate design allows Disen-CGCN to surpass MB-CGCN in recommendation performance, thereby demonstrating the efficacy of our model.

\begin{figure}[t]
\centering
 \subfloat[Preferences for item $i_{1}$ by user $u_{1}$]
 {\includegraphics[width=0.45\linewidth]{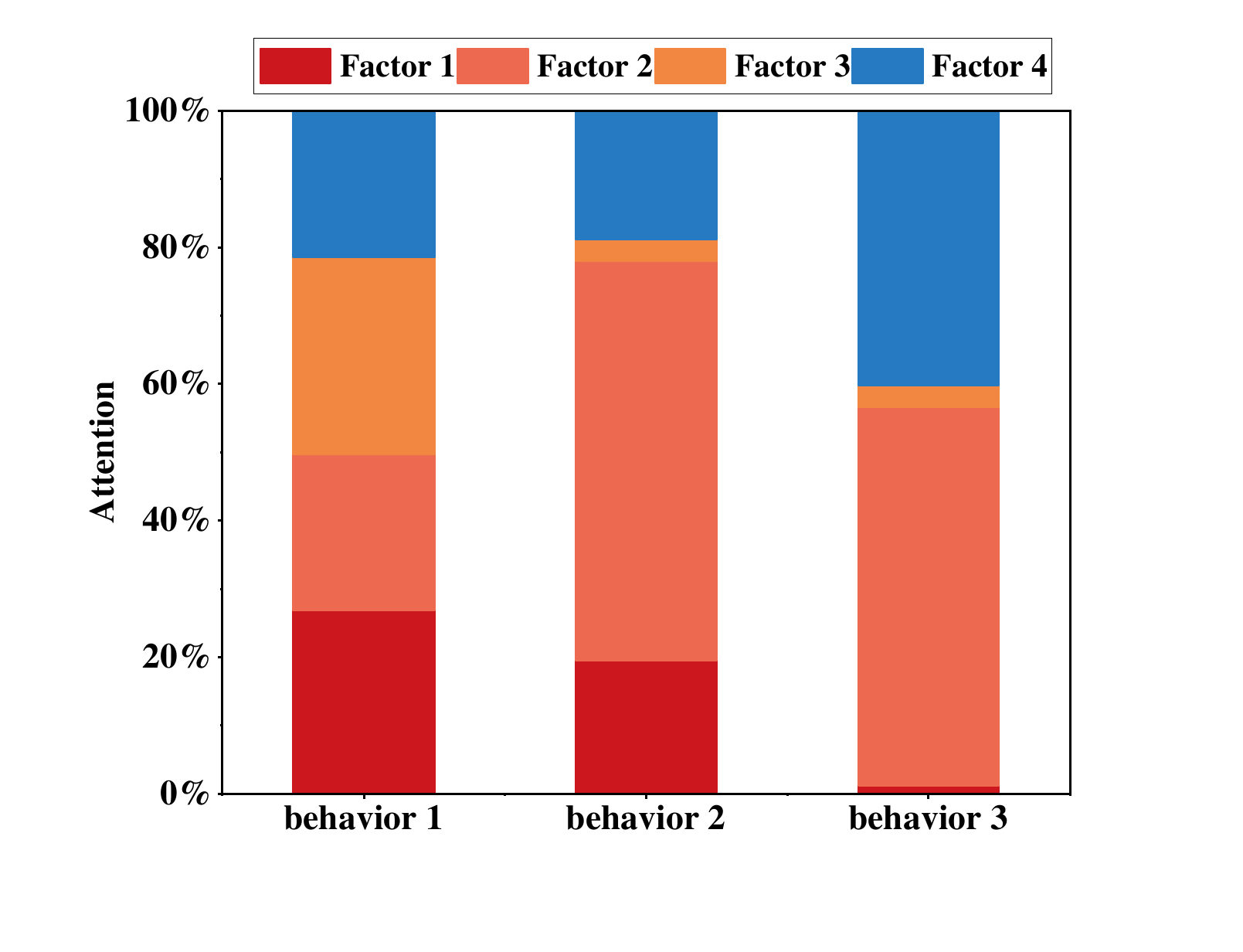}}
 \subfloat[Preferences for item $i_{2}$ by user $u_{1}$]
 {\includegraphics[width=0.45\linewidth]{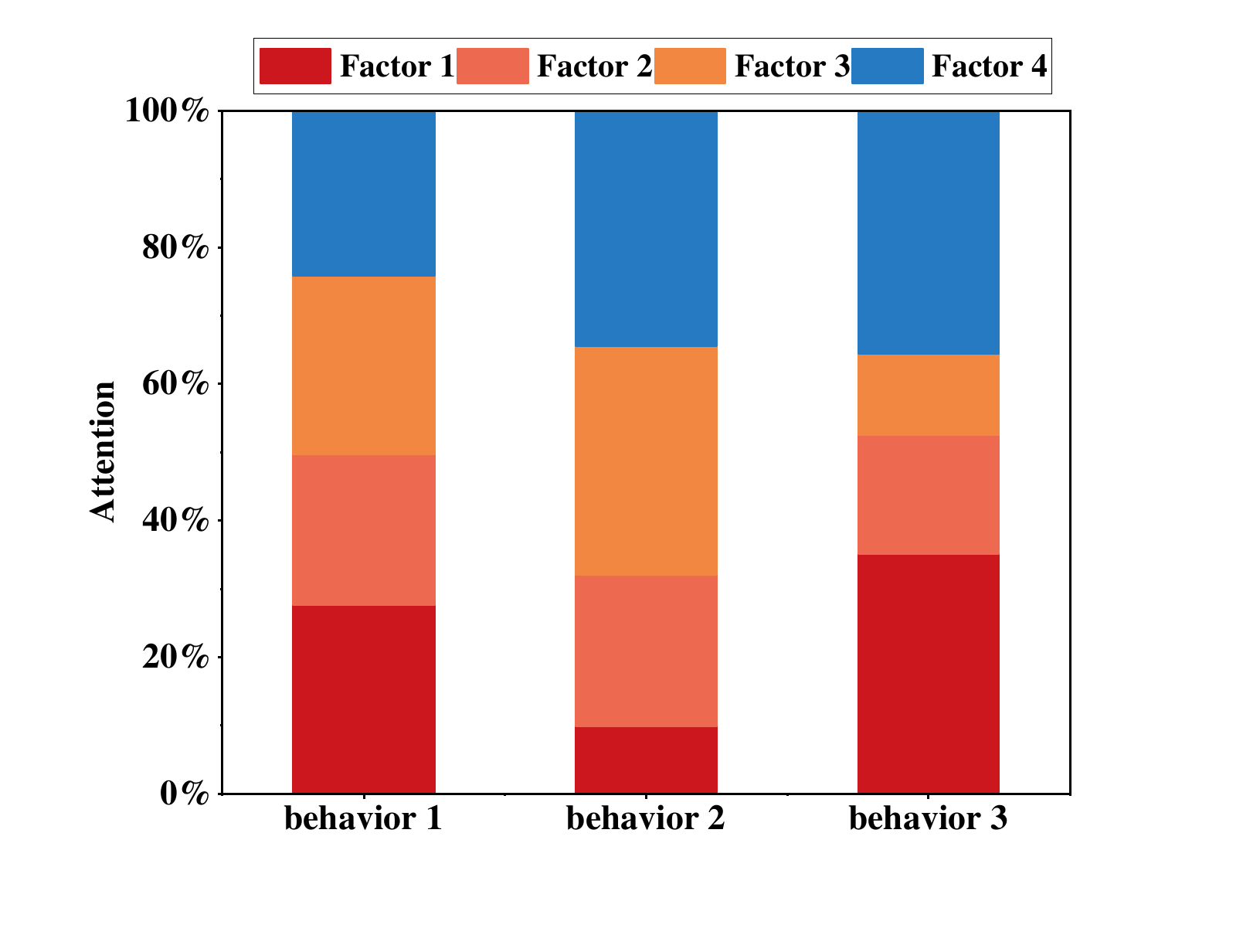}}\\
 \subfloat[Preferences for item $i_{3}$ by user $u_{2}$]
 {\includegraphics[width=0.45\linewidth]{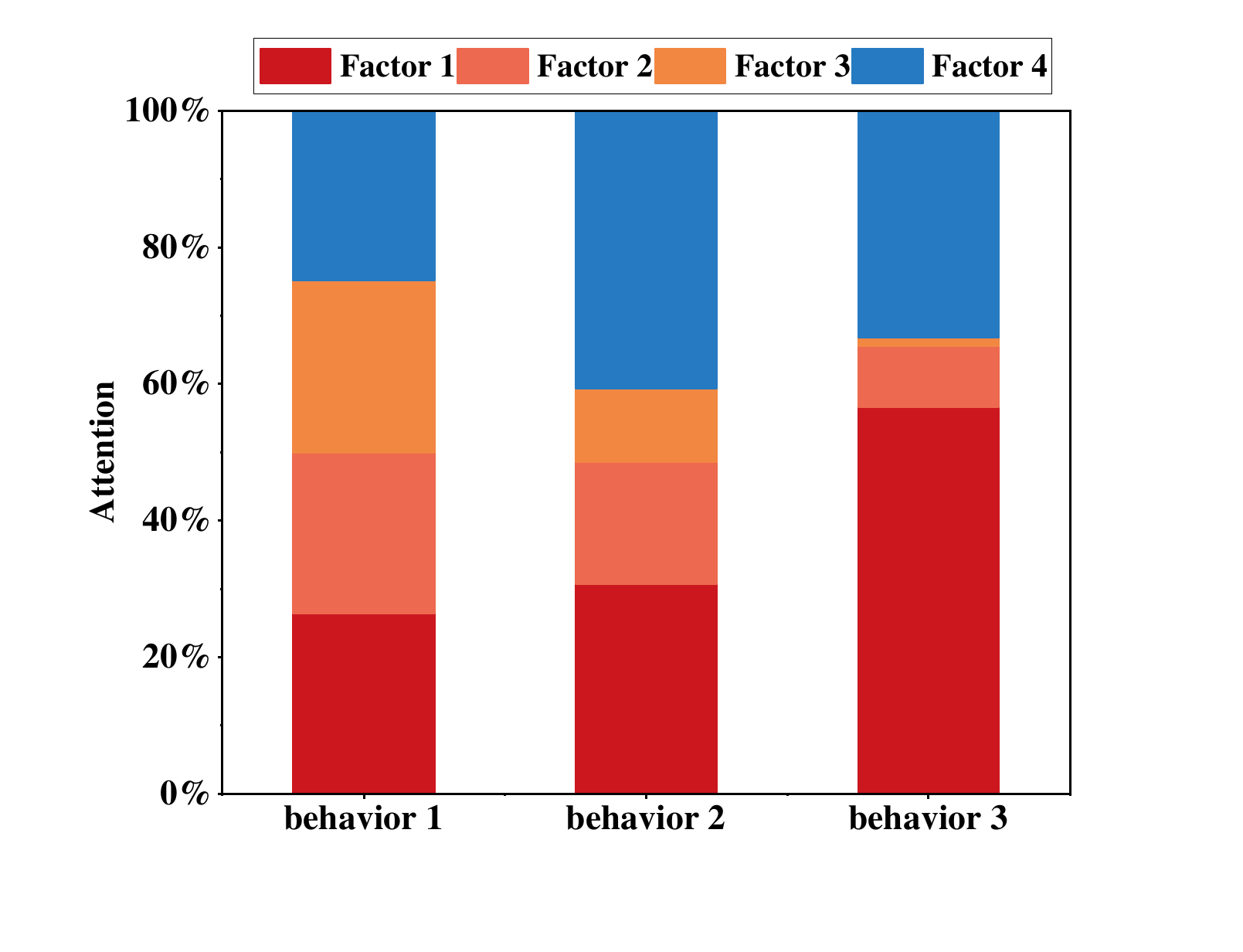}}
 \subfloat[Preferences for item $i_{3}$ by user $u_{3}$]
 {\includegraphics[width=0.45\linewidth]{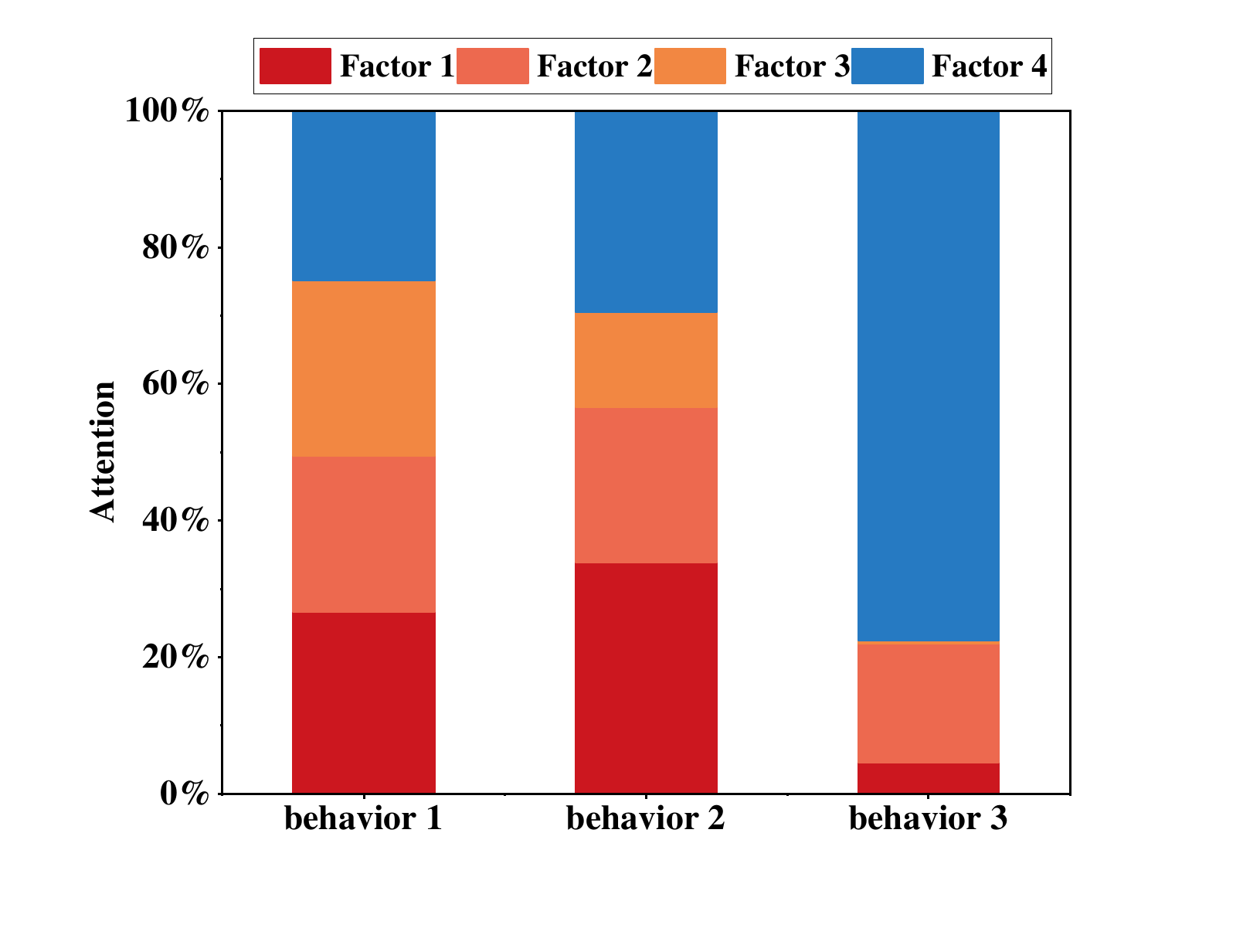}}
 %\subfloat[True location attribute(69)]
 %{\includegraphics[width=0.35\linewidth]{Fig/bc_loc_r.png}}
 \caption{Visualize the different preferences of users in different behaviors.}
 \label{fig:visual_attention}
\end{figure}

\subsection{ User Preference Analysis (RQ2)}
Our Disen-CGCN model is designed to capture users' preferences for different factors of items across various behaviors. We employ an attention mechanism (refer to Eq.~\ref{equation:attention}) to calculate the weights of users' attention towards different item factors in each behavior. To validate and vividly explain the motivation, we use two sets of visualizations based on the Beibei dataset. Firstly, we select a user and two items to examine the same user's varying preferences for different items across behaviors. Secondly, we choose two users and one item to explore how different users perceive the same item differently in various behaviors. 

In these visualizations, the horizontal axis represents different behaviors following a cascade sequence (view > cart > buy), while the vertical axis shows the weights of different factors, divided into $K$ blocks summing to 1. Each behavior is depicted in various colors, representing different factors, with the size of each color indicating the user's attention level to that factor.

Typically, user behavior in the initial phase is driven by superficial factors such as visual or cost, and as the behavior progresses, preference factors gradually shift to detailed attributes (e.g., comfort, quality, functionality, etc.) to guide it. From Figures ~\ref{fig:visual_attention}(a) and ~\ref{fig:visual_attention}(b), we can observe that users' preferences for different items are similar yet different across behaviors. This suggests that users' intrinsic preferences remain consistent across items and behaviors, and this intrinsic preference may be due to reasons such as users' preference for shopping for items with high utility. For example, in Factor 4 (where the utility of the item is indicated by the blue block), the user's preference for both items increases gradually as the behavior progresses. However, preferences for other factors may vary depending on an individual's need for the item. For example, for item $i_1$, users may pay more attention to comfort (factor 2), yet for item $i_2$, users pay attention to price (factor 1). Comparing Figures ~\ref{fig:visual_attention}(c) and ~\ref{fig:visual_attention}(d), we note that different users show similar but unique preferences for the same item across behaviors. This implies that certain intrinsic factors of the item remain constant across behaviors, such as price, scent, or the appearance of an expensive perfume. As the behavior evolves, attention to appearance (Factor 3) may decline consistently across all users, while other factors may vary according to the user's personal preferences. This example also emphasizes the importance of the personalized feature transformation we designed, where the shopping habits of  $u_2$ users are dominated by the price factor (factor 1), yet  $u_3$ users are more concerned with utility (factor 4). This shows that even for the same items, behavioral preferences vary from user to user.

An interesting pattern emerges from these visualizations: users' preferences for different factors tend to be similar in early behaviors (e.g., Behavior 1), possibly because users initially seek to understand all the factors of an item and compare it to similar items. This is consistent with typical shopping behavior. However, as the behavior evolves, the user's preferences become more focused on certain factors that will ultimately influence the user's final decision-making process for the item.

\subsection{Ablation Study (RQ3)}
In this section, we conduct comprehensive ablation experiments to validate the key components of the Disen-CGCN model. Our model advances beyond the MB-CGCN framework by specifically accounting for distinct user preferences across different factors in each behavior.  To understand the impact of various elements, we compare our model with three variants and MB-CGCN, focusing on the contributions of the disentangled representation technique, the attention mechanism, and the personalized feature transformation. The three variants used in the experiment are:

\begin{itemize}
    \item \textbf{w/o. A\&T  (without Attention and Personalized Transformation)}: This variant removes both the attention mechanism and personalized feature transformation from our Disen-CGCN model. It can be also regarded as a model that incorporates the disentangled representation technique into MB-CGCN, leading to the separation of user and item embeddings. It is important to note that during feature transformation between behaviors, we apply a shared feature transformation for different factors.

    \item \textbf{w/o. A (without Attention)}:  In this version, the attention mechanism is removed from Disen-CGCN. This implies that the user’s preference for items remains constant across different behaviors.

    \item \textbf{w/o. T  (without Personalized Transformation)}: This variant excludes the personalized feature transformation from Disen-CGCN. Consequently, similar to MB-CGCN, it employs a shared feature transformation for different factors of users and items between behaviors.
\end{itemize}

The three variants used in the experiment are designed to isolate and evaluate the individual contributions of these key features. By comparing the performance of each variant with the full Disen-CGCN model, we aim to quantify the impact of the disentangled representation technique, the attention mechanism, and the personalized feature transformation on the overall effectiveness of our model.  It is important to note that when all three modules are excluded from our Disen-CGCN model, it essentially reverts to the MB-CGCN model. Therefore, we include the performance of MB-CGCN in our results for a comprehensive analysis.  This comparative comparison is intended to shed light on the significance of each component, illustrating how they collectively enhance the model's capability to accurately capture and predict user preferences across various behaviors. The results of this experiment are presented in Table~\ref{Ablation_Study}.

\subsubsection{\textbf{Effects of disentanglement.}}
To validate the effectiveness of disentangled representation learning in our Disen-CGCN model, we conducted a comparative analysis with the MB-CGCN model and the variant \textit{w/o. A\&T}.  In these experiments, disentanglement was applied exclusively to the features in the first behavior. This involved separating the inputs of the first behavior and the embeddings learned therein. We ensured the independence of factors for both user and item in this first behavior, following which the embeddings of these distinct factors were transformed into separate features.

 The incorporation of disentangled representation learning alone led to an average improvement of 1.15\% and 1.68\% across all metrics when compared to the backbone model (MB-CGCN) on the Beibei and Tmall datasets, respectively. These findings underscore the advantage of separating various entangled factors in user and item embeddings to more accurately model user preferences for items. Consequently, this experiment substantiates the efficacy of our Disen-CGCN model in leveraging disentangled representation learning.

\begin{table}[]
\centering
\normalsize
\caption{Ablation study of our proposed Disen-CGCN method over two datasets.}
\label{Ablation_Study}
\begin{tabular}{cl|cc|cc|cc}
\hline
\multicolumn{1}{l}{\textit{\textbf{}}} &
  \textbf{Model} &
  \textbf{R@10} &
  \textbf{N@10} &
  \textbf{R@20} &
  \textbf{N@20} &
  \textbf{R@50} &
  \textbf{N@50} \\ \hline\hline
\multirow{5}{*}{\textit{\textbf{Beibei}}} & {MB-CGCN} & 0.0580 & 0.0288 & 0.0995 & 0.0392 & 0.1933 & 0.0577 \\ 
                                          & {w/o. A\&T}     & 0.0595 & 0.0289 & 0.1012 & 0.0394 & 0.1956 & 0.0579 \\ 
                                          & {w/o. A}   & 0.0606 & 0.0306 & 0.1030 & 0.0412 & 0.2020 & 0.0607 \\ 
                                          & {w/o. T}   & 0.0591 & 0.0293 & 0.1015 & 0.0399 & 0.1976 & 0.0588 \\ 
 &
  \textbf{Disen-CGCN} &
  \textbf{0.0620} &
  \textbf{0.0314} &
  \textbf{0.1055} &
  \textbf{0.0423} &
  \textbf{0.2044} &
  \textbf{0.0617} \\ \hline
\multirow{5}{*}{\textit{\textbf{Tmall}}}  & {MB-CGCN} & 0.1317 & 0.0686 & 0.2007 & 0.0859 & 0.3226 & 0.1101 \\ 
                                          & {w/o. A\&T}     & 0.1339 & 0.0695 & 0.2040 & 0.0872 & 0.3296 & 0.1120 \\ 
                                          & {w/o. A}   & 0.1438 & 0.0746 & 0.2151 & 0.0926 & 0.3415 & 0.1175 \\ 
                                          & {w/o. T}   & 0.1376 & 0.0708 & 0.2064 & 0.0881 & 0.3357 & 0.1137 \\ 
 &
  \textbf{Disen-CGCN} &
  { \textbf{0.1469}} &
  { \textbf{0.0761}} &
  { \textbf{0.2172}} &
  { \textbf{0.0938}} &
  { \textbf{0.3430}} &
  { \textbf{0.1186}} \\ \hline
\end{tabular}
\end{table}

\subsubsection{\textbf{Effects of attention mechanisms.}}
The attention mechanism in the Disen-CGCN model plays a crucial role in distinguishing users' varying preferences for items across different behaviors. To validate its effectiveness, we initially compare the performances of \textit{w/o. A\&T} and \textit{w/o. T}, where \textit{w/o. T }represents the extension of \textit{w/o. A\&T} with the addition of the attention mechanism. This addition shifts the model from capturing users' uniform preferences across behaviors to recognizing diverse preferences. This comparison results in an average improvement of 0.85\% and 1.74\% across all metrics on the Beibei and Tmall datasets, respectively. Further, by comparing \textit{w/o. A} (the variant without the attention mechanism) with the full Disen-CGCN model, we observe a decrease in performance when the attention mechanism is removed. This performance gap between \textit{w/o. A} and Disen-CGCN unequivocally demonstrates the effectiveness of incorporating the attention mechanism in our model. These two sets of experimental comparisons collectively affirm that the attention mechanism significantly enhances the Disen-CGCN model's ability to accurately capture user preferences in various behaviors.

\subsubsection{\textbf{Effects of personalized feature transformation.}}
%We are committed to modeling more fine-grained multi-behavior recommender systems, so we model personalized feature transformations for users and items separately for information transfer between behaviors. To demonstrate that personalized feature transformations are effective, we compare two sets of experiments as we did for exploring the attention mechanism. First, we compare w.D and w/o.A. w/o.A is a further improvement of w.D, which replaces shared feature transformations with personalized feature transformations for each user and each item separately. The results show an average improvement of 3.71\% and 5.9\% for all metrics on the Beibei and Tmall datasets, respectively. Significant improvement can be seen using personalized feature transformation. We also compare the performance of w/o.T and Disen-CGCN, and the model performance decreases significantly after removing the personalized feature transformation. This shows that the personalized feature transformation module plays a very important role in improving the performance of our model.
We aim to develop more fine-grained multi-behavior recommender systems. Therefore, in our Disen-CGCN model, we implement personalized feature transformations for users and items to facilitate information transfer between behaviors. To demonstrate the effectiveness of these personalized transformations, we conducted two sets of comparative experiments, similar to our approach in assessing the attention mechanism.

Initially, we compared \textit{w/o. A\&T} with \textit{w/o. A}, where \textit{w/o. A} represents an advancement over \textit{w/o. A\&T} by incorporating personalized feature transformations for each user and item, in contrast to shared feature transformations. This modification resulted in an average performance enhancement of 3.71\% and 5.9\% across all metrics on the Beibei and Tmall datasets, respectively. This significant improvement underscores the efficacy of personalized feature transformation.
Furthermore, we compared the performance of \textit{w/o. T} and the full Disen-CGCN model. Removing the personalized feature transformation from Disen-CGCN led to a notable decrease in model performance, highlighting the critical role this module plays in our model's improved functionality. These comparisons collectively validate that personalized feature transformation is a key factor in enhancing the performance of our Disen-CGCN model.

% \usepackage{multirow}
\begin{table}[]
\centering
\normalsize
\caption{Effects of different meta-knowledge learning methods.}
\label{meta knowledge select}
\begin{tabular}{cl|cc|cc|cc}
\hline
\multicolumn{1}{l}{\textit{\textbf{}}} &
  \textbf{Model} &
  \textbf{R@10} &
  \textbf{N@10} &
  \textbf{R@20} &
  \textbf{N@20} &
  \textbf{R@50} &
  \textbf{N@50} \\ \hline\hline
\multirow{2}{*}{\textit{\textbf{Beibei}}} &
  w. post &
  0.0593 &
  0.0297 &
  0.1036 &
  0.0408 &
  0.1988 &
  0.0595 \\ 
 &
  \textbf{Ours} &
  \textbf{0.0620} &
  \textbf{0.0314} &
  \textbf{0.1055} &
  \textbf{0.0423} &
  \textbf{0.2044} &
  \textbf{0.0617} \\ \hline
\multirow{2}{*}{\textit{\textbf{Tmall}}} &
  w. post &
  0.1450 &
  0.0746 &
  0.2159 &
  0.0925 &
  0.3424 &
  0.1175 \\ 
 &
  \textbf{Ours} &
  \textbf{0.1469} &
  \textbf{0.0761} &
  \textbf{0.2172} &
  \textbf{0.0938} &
  \textbf{0.3430} &
  \textbf{0.1186} \\ \hline
\end{tabular}
\end{table}

% % Please add the following required packages to your document preamble:
% % \usepackage{multirow}
% \begin{table}[]
% \centering
% \normalsize
% \caption{Effects of different meta-knowledge in Disen-CGCN.}
% \label{meta knowledge select}
% \begin{tabular}{cl|cc|cc|cc}
% \hline
% \multicolumn{1}{l}{\textit{\textbf{}}} &
%   \textbf{Model} &
%   \textbf{R@10} &
%   \textbf{N@10} &
%   \textbf{R@20} &
%   \textbf{N@20} &
%   \textbf{R@50} &
%   \textbf{N@50} \\ \hline\hline
% \multirow{2}{*}{\textit{\textbf{Beibei}}} &
%   w.ALL &
%   5.93 &
%   2.97 &
%   10.36 &
%   4.08 &
%   19.88 &
%   5.95 \\ 
%  &
%   \textbf{w.First(Ours)} &
%   \textbf{6.20} &
%   \textbf{3.14} &
%   \textbf{10.55} &
%   \textbf{4.23} &
%   \textbf{20.44} &
%   \textbf{6.17} \\ \hline
% \multirow{2}{*}{\textit{\textbf{Tmall}}} &
%   w.ALL &
%   14.50 &
%   7.46 &
%   21.59 &
%   9.25 &
%   34.24 &
%   11.75 \\ 
%  &
%   \textbf{w.First(Ours)} &
%   \textbf{14.69} &
%   \textbf{7.61} &
%   \textbf{21.72} &
%   \textbf{9.38} &
%   \textbf{34.30} &
%   \textbf{11.86} \\ \hline
% \end{tabular}
% \end{table}

\subsubsection{\textbf{Effects of meta-knowledge learning methods.}} \label{section:mks}
%In Equation~\ref{equation:T}, we opted to use both the embeddings of users and items themselves and the aggregation of their first-order neighbors' embeddings learned from the previous behavior during graph convolution as meta-knowledge. This approach differs from the methods used in \cite{xia2021graph,chen2023heterogeneous}, which typically employ the nodes and their neighbors' information post-graph convolution. To assess the effectiveness of these differing strategies, we compared their performances as shown in Table~\ref{meta knowledge select}. The method \textit{w. post} represents the use of neighbor information after convolution. The results indicate that our model outperforms \textit{w. post} across all metrics in every scenario. This superiority demonstrates the advantages of our strategy on distilling meta-knowledge in multi-behaviors. Our method considers knowledge from both the current and previous behavior like MB-CGCN~\cite{yan2023cascading} in passing the information to the next behavior, which has been demonstrated to be effective.  Notably, in the Beibei dataset, where the optimal number of graph convolution layers for the model is $\{3,4,3\}$ layers, the performance using our method is significantly better than  that of \textit{w. post} as in previous work.
In our approach, detailed in Equation~\ref{equation:T}, we integrate user and item embeddings with aggregated embeddings of their first-order neighbors, which are derived from previous behaviors during graph convolution. This method contrasts with the ones employed in \cite{xia2021graph,chen2023heterogeneous}, where node and neighbor information is primarily used after the graph convolution process. To evaluate the effectiveness of these differing methodologies, we compared their performances as presented in Table~\ref{meta knowledge select}. In this comparison, the \textit{w. post} method signifies the utilization of neighbor information post-convolution. Our results demonstrate that our model consistently surpasses the \textit{w. post} method across all metrics in various scenarios, highlighting the efficacy of our strategy in extracting meta-knowledge from multi-behavioral data. This superiority is further validated by the similarities between our method and the MB-CGCN approach described in \cite{yan2023cascading}. Both approaches emphasize the critical role of incorporating information from both current and previous behaviors in transmitting data to subsequent behaviors.

\subsection{Impact of hyperparameters (RQ4)}
In this section, we study the influence of three critical hyperparameters on our model across two datasets: the number of factors, the number of GCN layers, and the attention coefficient (i.e., $\rho$ in Eq.~\ref{equation:attention}). We report the results for top-$n$ ($n$=\{20\}) and omit the results for top-$n$ ($n$=\{10, 50\}) as they show exactly the same trend.

\begin{figure}[t]
\centering
 \hspace{-0.4 cm}
 {\includegraphics[width=0.28\linewidth]{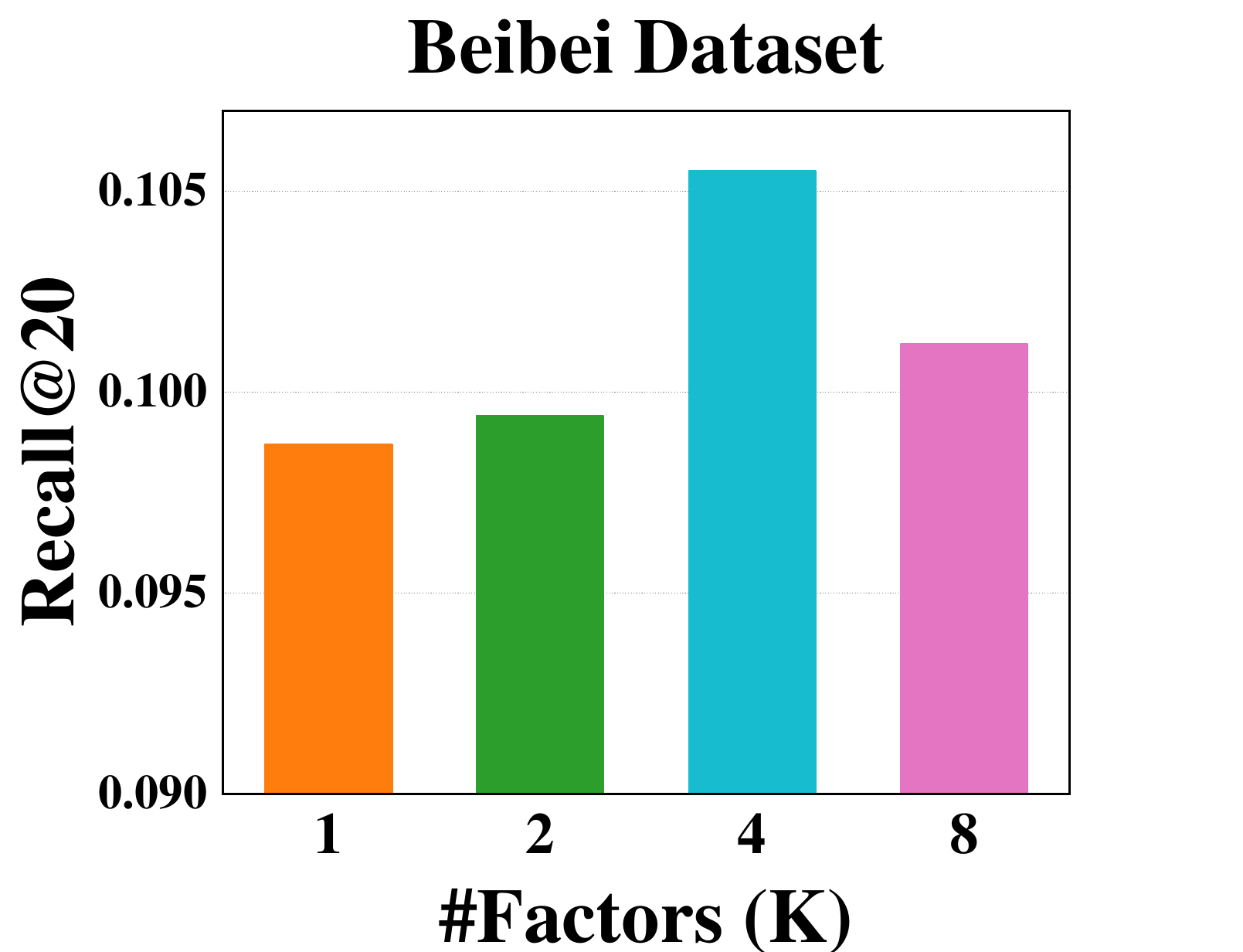}} \hspace{-0.6cm}
 {\includegraphics[width=0.28\linewidth]{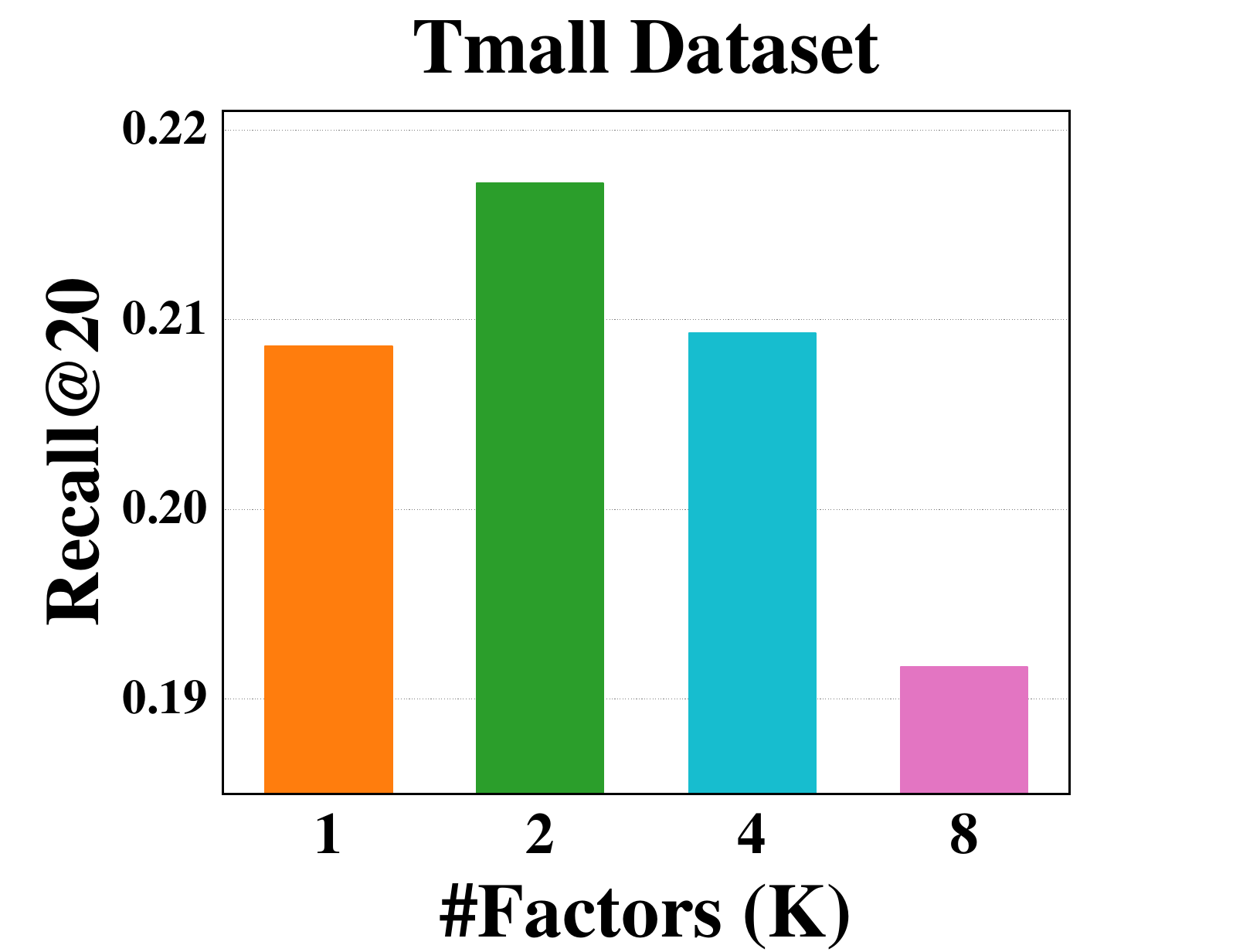}}\hspace{-0.6cm}
 {\includegraphics[width=0.28\linewidth]{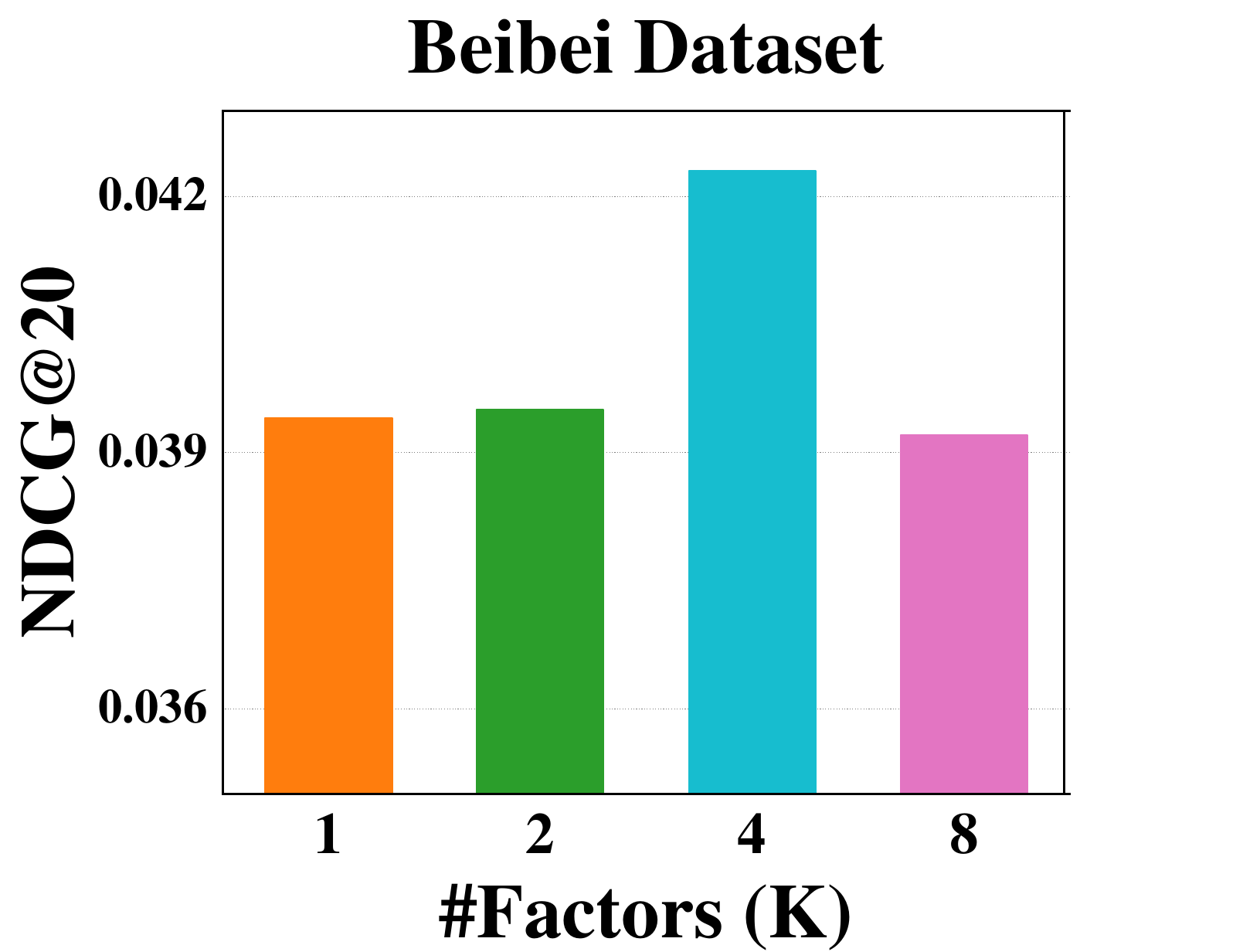}}\hspace{-0.6cm}
 {\includegraphics[width=0.28\linewidth]{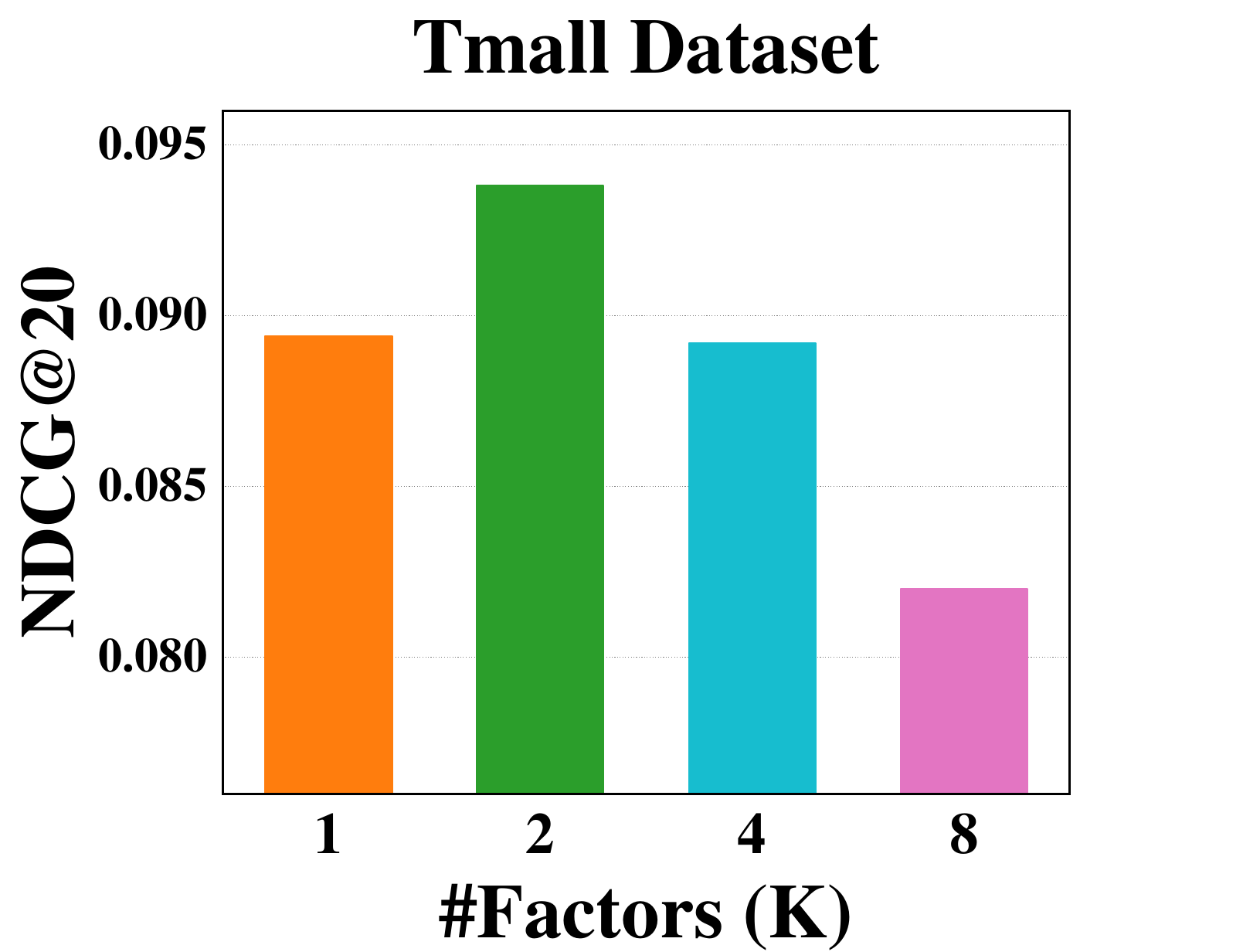}}\hspace{-0.6cm}
% \vspace{8pt}
 \caption{Impact of the number of factors (K).}
 \vspace{0pt}
 \label{fig:factors}
\end{figure}

\subsubsection{\textbf{Impact of factor number.}}

We explored how varying the number of factors ($K$) in disentangled representation learning affects the model, keeping the user and item embedding size constant at 
$d=64$.  We experimented with different factor counts in the set $\{1,2,4,8\}$, where each factor's embedding size is $\frac{d}{K}$.
Figure~\ref{fig:factors} illustrates that the performance on both datasets generally improves with an increasing number of factors but starts to decline after reaching a certain point. This peak represents the optimal number of factors for each dataset. A low number of factors ($K=1$) yields suboptimal performance, as it fails to capture fine-grained user preferences. Conversely, at higher counts ($K=4$ or 8), the reduced embedding size per factor limits the expressiveness of each factor's representation. Notably, the optimal number of factors varies between datasets, 4 for the Beibei dataset and 2 for the Tmall dataset, reflecting differing user preference complexities in various scenarios.

\begin{figure}[t]
\centering
 \hspace{-0.4 cm}
 {\includegraphics[width=0.28\linewidth]{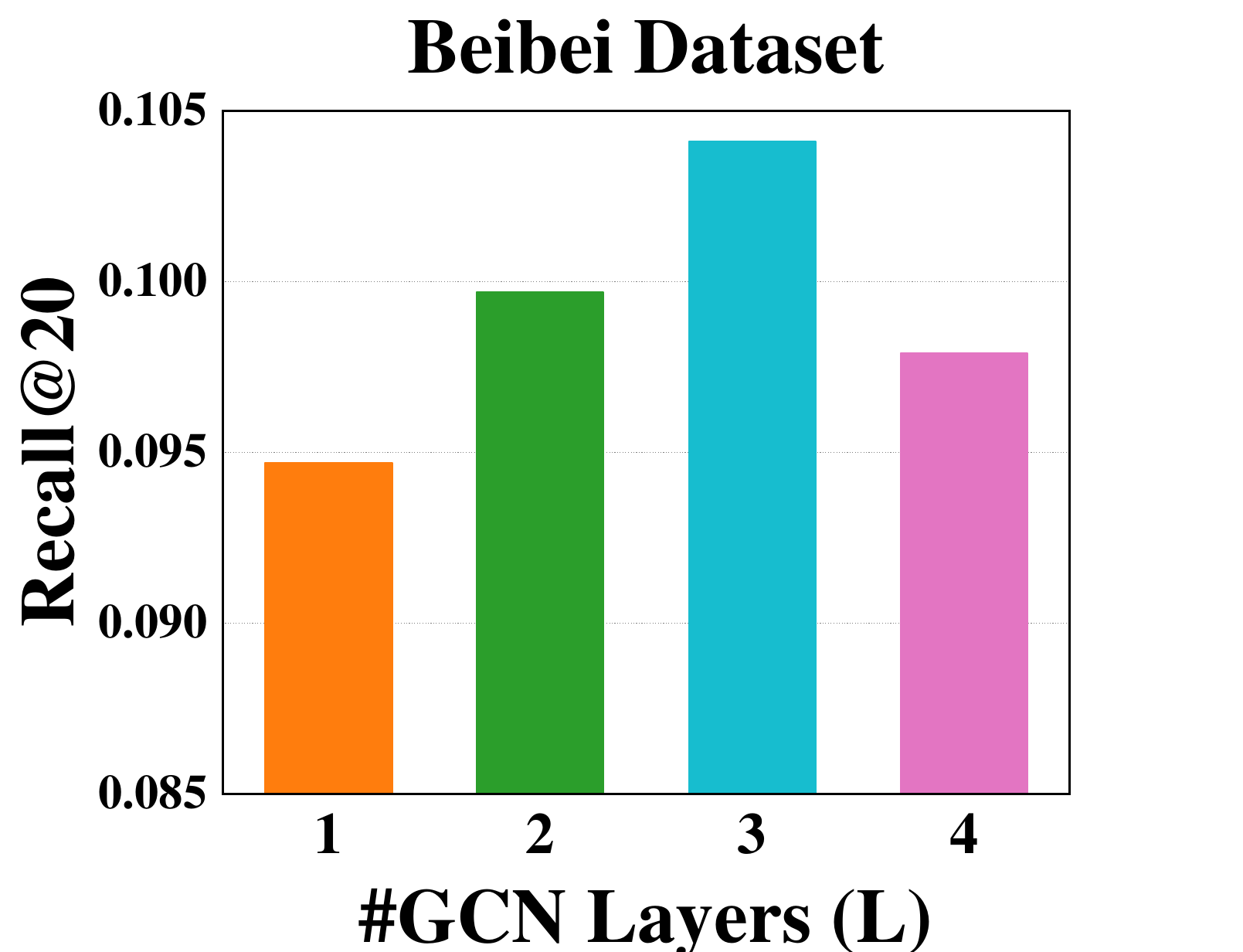}}  \hspace{-0.6cm}
 {\includegraphics[width=0.28\linewidth]{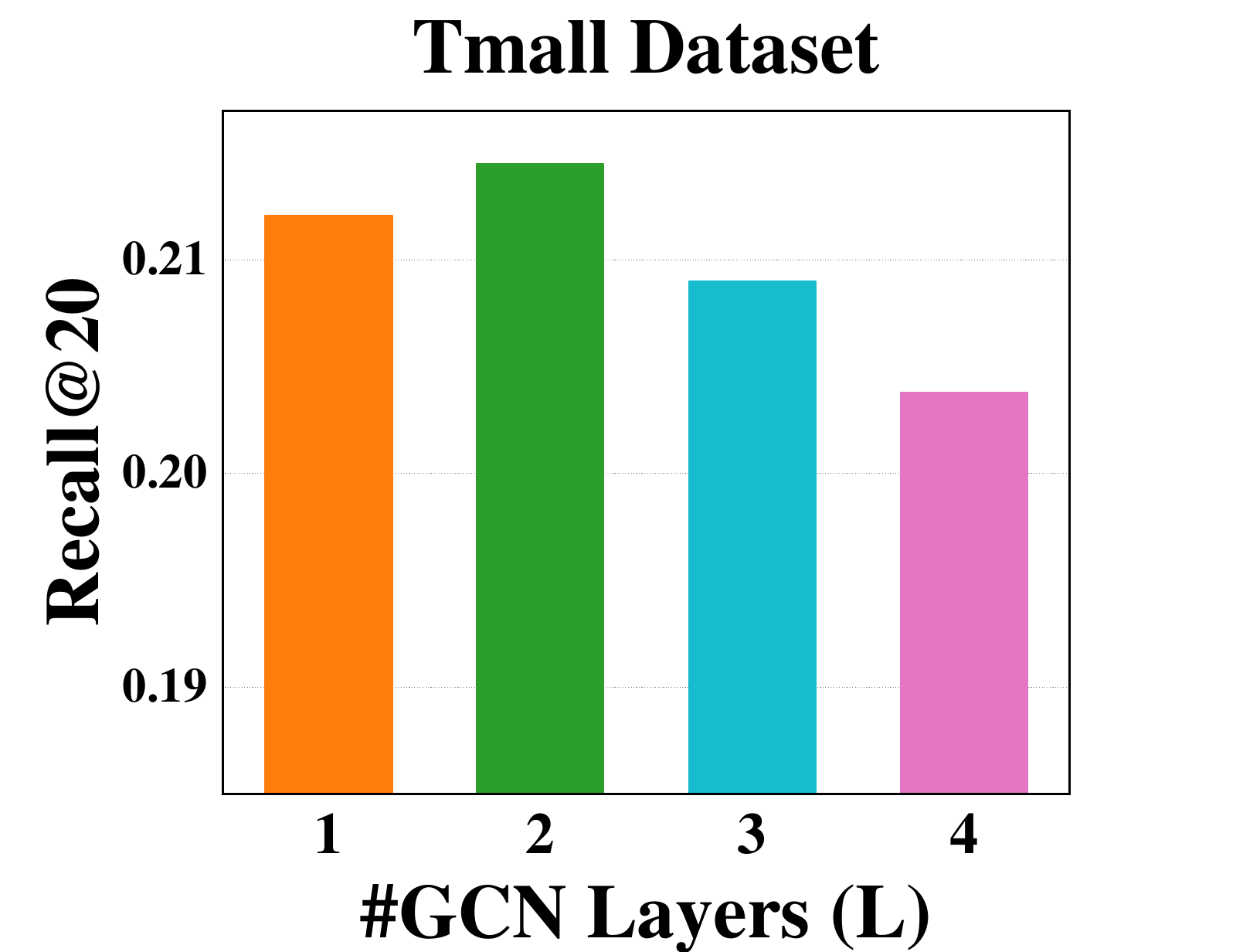}}  \hspace{-0.6cm}
 {\includegraphics[width=0.28\linewidth]{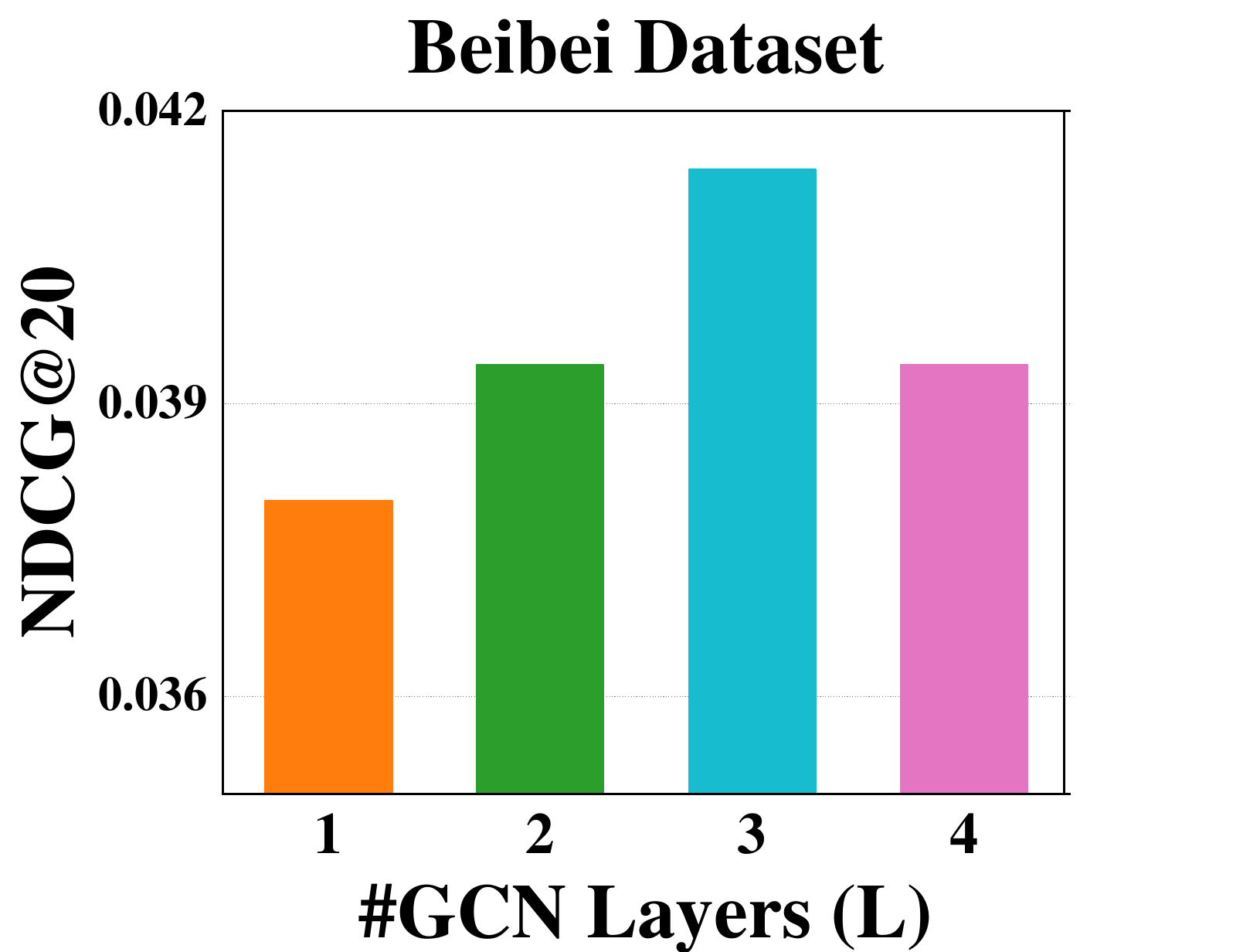}} \hspace{-0.6cm}
 {\includegraphics[width=0.28\linewidth]{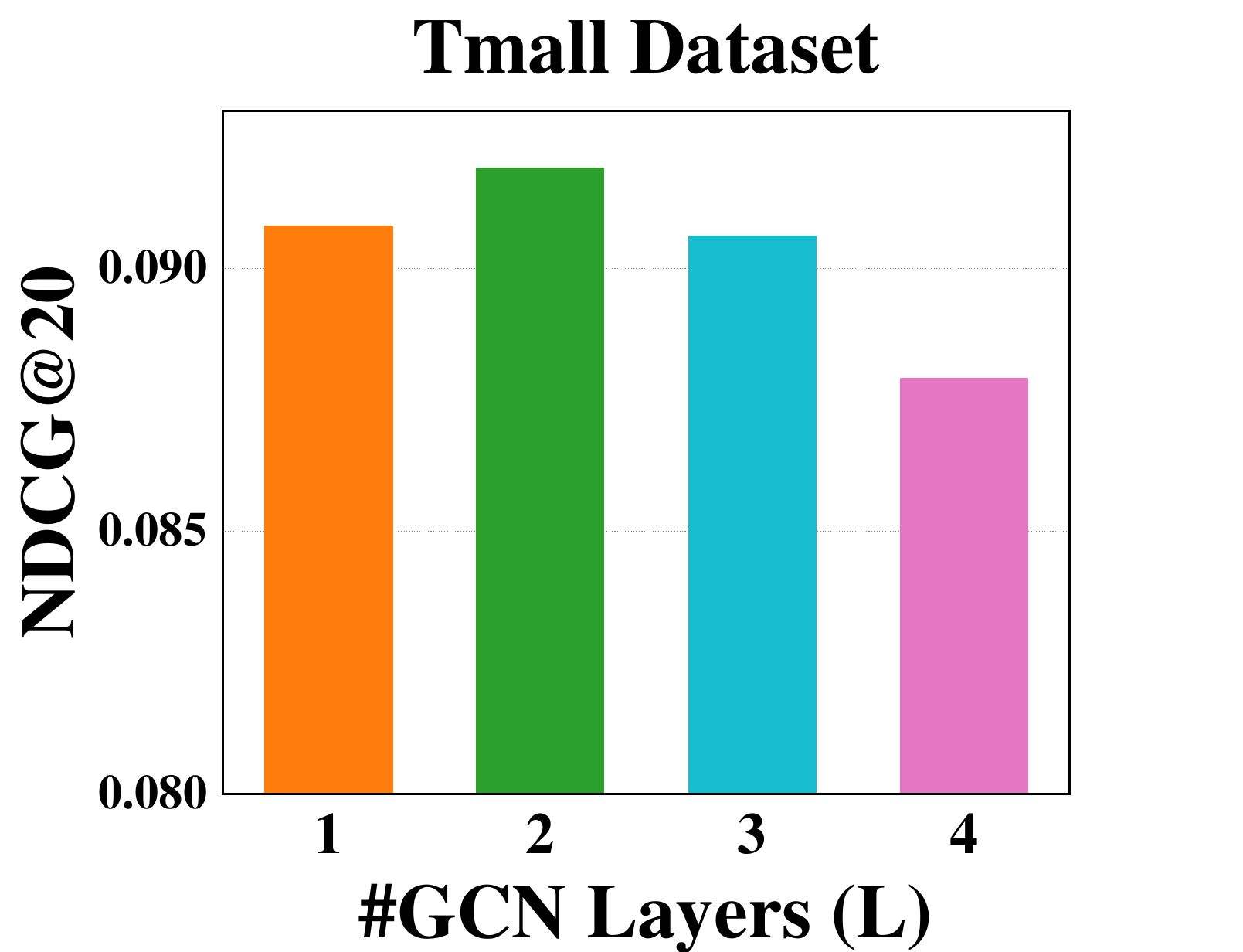}} \hspace{-0.6cm}
 \caption{Impact of the number of GCN layers (L).}
 \label{fig:gcn}
\end{figure}

% \begin{figure}[t]
% \centering
%  \hspace{0.0cm}
%  %\subfloat[Age (ML-1M)]
%  {\includegraphics[width=0.45\linewidth]{Fig/DBRLayer_en.pdf}}
%  %\subfloat[Occupation (ML-1M)]
%  {\includegraphics[width=0.45\linewidth]{Fig/DBNLayer_en.pdf}}
%  {\includegraphics[width=0.45\linewidth]{Fig/DTRLayer_en.pdf}}
%  {\includegraphics[width=0.45\linewidth]{Fig/DTNLayer_en.pdf}}
%  %{\includegraphics[width=0.35\linewidth]{Fig/bc_loc_r.png}}
%  \vspace{8pt}
%  \caption{Impact of the GCN layers number.}
%  \vspace{0pt}
%  \label{fig:gcn}
% \end{figure}

\subsubsection{\textbf{Impact of the number of GCN layers.}}
We also investigated the impact of the number of GCN layers, standardizing their count across all behaviors for comparative purposes. As shown in Figure~\ref{fig:gcn}, an increase in the number of GCN layers initially enhances performance in both datasets, attributable to our use of LightGCN as the backbone network for embedding learning. More GCN layers enable better utilization of higher-order neighbor information in the user-item bipartite graph.
However, there is a limit to this improvement. Performance begins to deteriorate when exceeding 4 layers for the Beibei dataset and 3 layers for the Tmall dataset. This decrease is likely due to overfitting in both datasets, a common issue when GCN-based recommendation models performs excessive graph convolution operations..

\begin{figure}[t]
\centering
 \hspace{-0.4 cm}
 {\includegraphics[width=0.28\linewidth]{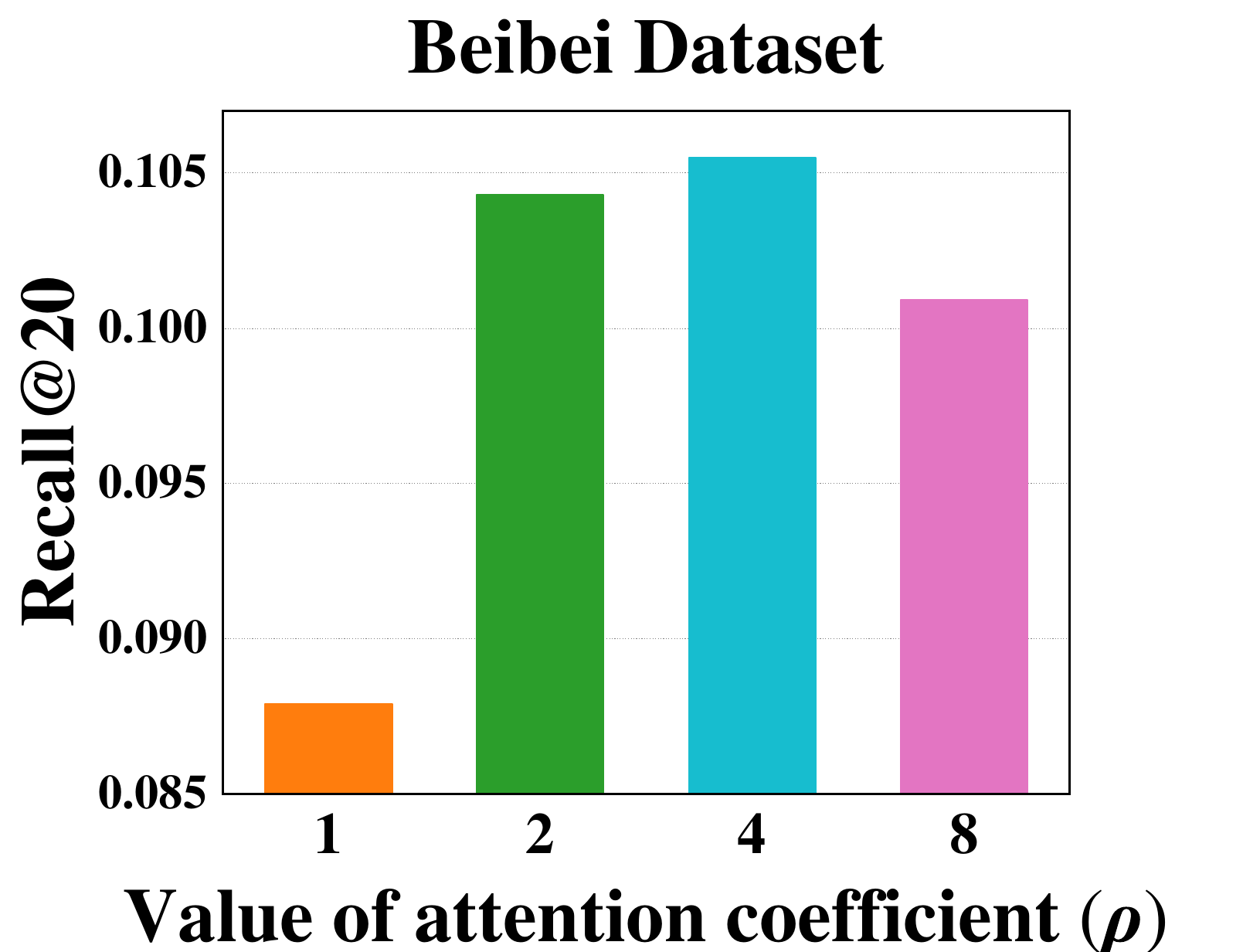}}  \hspace{-0.6cm}
 {\includegraphics[width=0.28\linewidth]{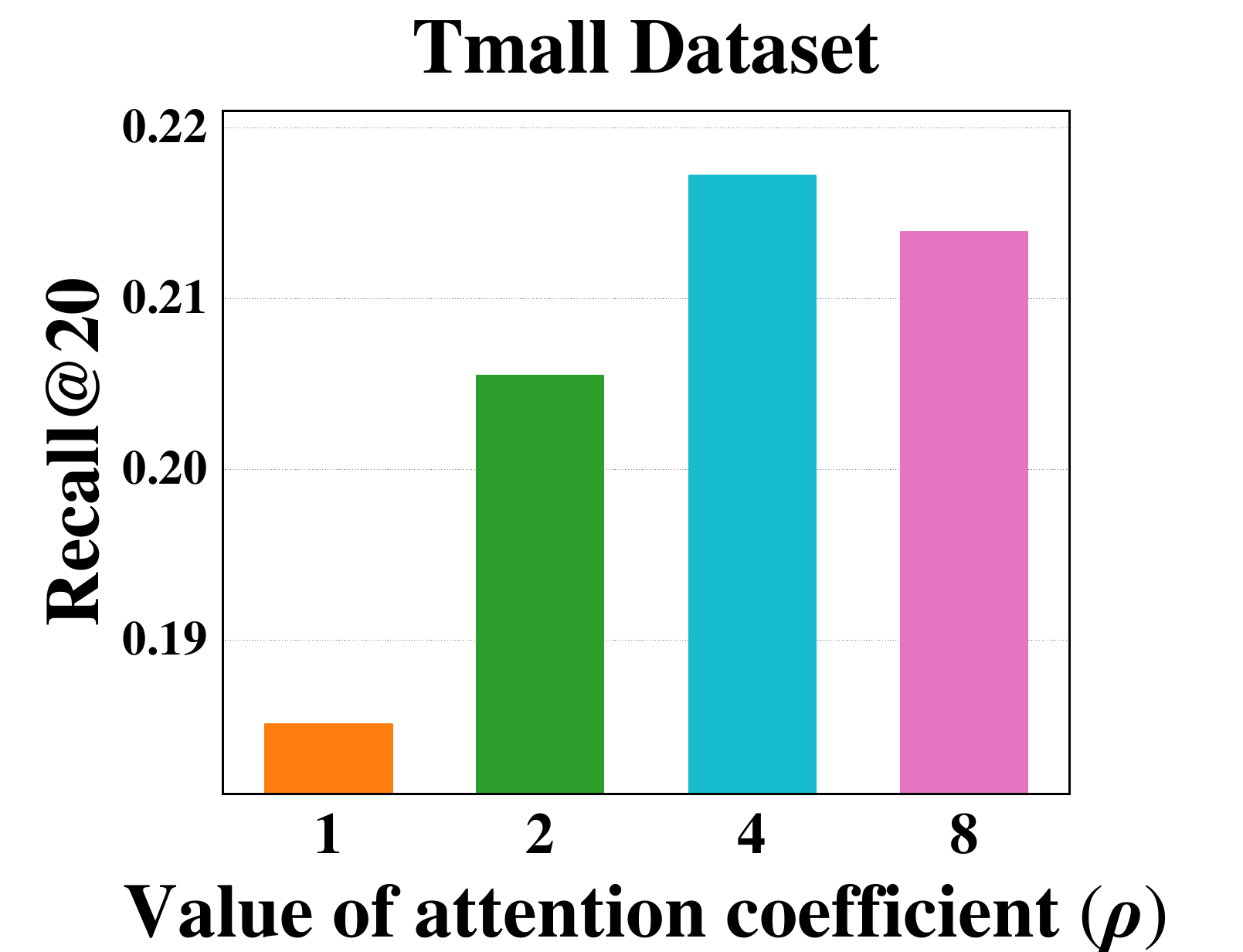}}  \hspace{-0.6cm}
 {\includegraphics[width=0.28\linewidth]{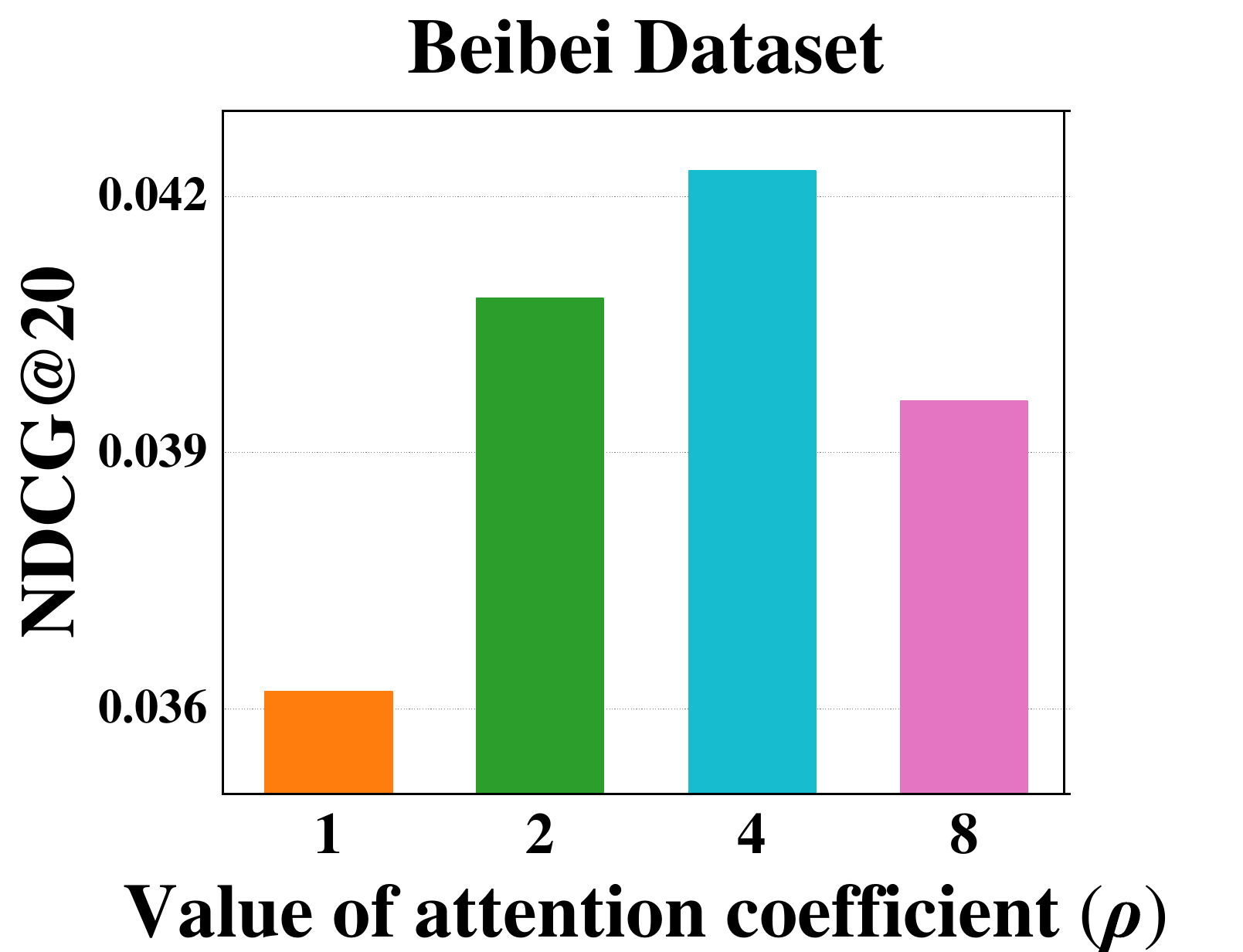}} \hspace{-0.6cm}
 {\includegraphics[width=0.28\linewidth]{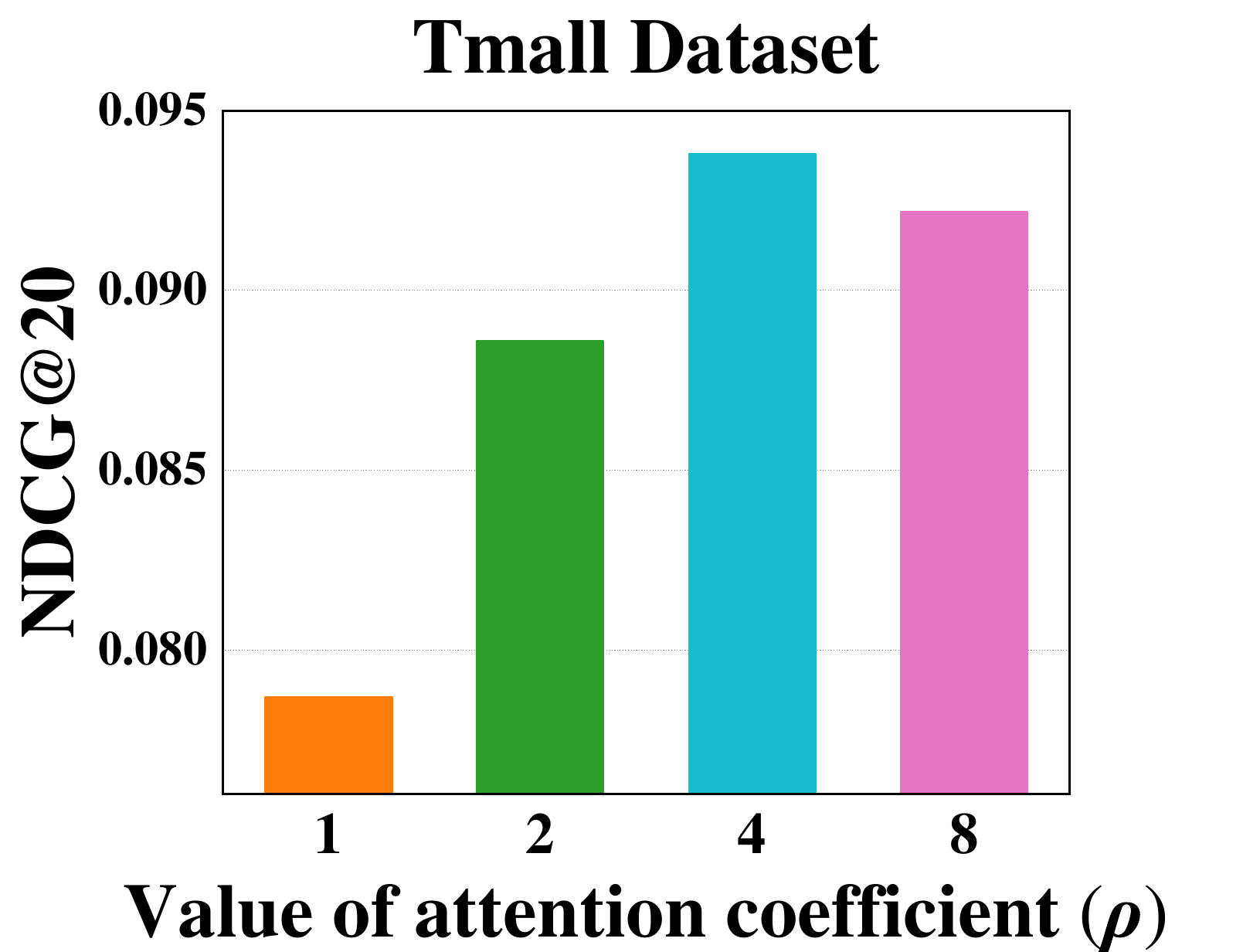}} \hspace{-0.6cm}
 \caption{Impact of attention coefficient ($\rho$).}
 \vspace{0pt}
 \label{fig:atte}
\end{figure}

% \begin{figure}[t]
% \centering
%  \hspace{0.0cm}
%  %\subfloat[Age (ML-1M)]
%  {\includegraphics[width=0.45\linewidth]{Fig/DBRA_en.pdf}}
%  %\subfloat[Occupation (ML-1M)]
%  {\includegraphics[width=0.45\linewidth]{Fig/DBNA_en.pdf}}
%  {\includegraphics[width=0.45\linewidth]{Fig/DTRA_en.pdf}}
%  {\includegraphics[width=0.45\linewidth]{Fig/DTNA_en.pdf}}
%  %\subfloat[True location attribute(69)]
%  %{\includegraphics[width=0.35\linewidth]{Fig/bc_loc_r.png}}
%  \vspace{8pt}
%  \caption{Impact of attention enlargement.}
%  \vspace{0pt}
%  \label{fig:atte}
% \end{figure}

\subsubsection{\textbf{Impact of attention coefficient.}}

Our model's design recognizes that the attention scores, which focus only on the user's $K$ block embeddings, might create a significant disparity in the magnitude of user and item embeddings. Following the insights from ~\cite{liu2019user}, we investigated how scaling up the attention scores—effectively enlarging the user's embeddings—affects the model's performance. The results, depicted in Figure~\ref{fig:atte}, provide valuable insights.

The model performs poorest when the enlargement factor $\rho = 1$, implying no amplification of the attention score. This indicates that a disparity in the magnitude of user and item embeddings can detrimentally affect the model's predictive capability. However, when the attentional enlargement factor $\rho$ is increased, we observe a substantial improvement in performance for both datasets. This suggests that aligning the magnitude of user and item embeddings is crucial for the model's efficacy.
Yet, there is a threshold to this improvement. As $\rho$ continues to rise, the model's performance begins to decline. This is attributed to the fact that excessively amplifying the user embeddings, without corresponding adjustments to item embeddings, leads to a new imbalance in their magnitudes.

In the Beibei dataset, the optimal attention coefficient aligns with the number of factors, both being 4. For the Tmall dataset, despite the optimal number of factors being 2, the ideal attention coefficient remains 4. This discrepancy may stem from the Tmall dataset's sparser nature, influencing the magnitude of learned user and item embeddings. Therefore, it appears that the attention coefficient should approximate the target number, corroborating the findings in \cite{liu2019user}.

\section{CONCLUSION} \label{conclusion}
%In this work, we propose a modeling fine-grained multi-behavior recommendation model, Disen-CGCN, which leverages the cascading relationships between behaviors. In each behavior, we employ disentangled representation techniques and attention mechanisms to capture different user preferences for items. Meanwhile, between behaviors, we design a meta-network to perform personalized feature transformation between users and items. Extensive experiments on two real datasets show that the Disen-CGCN model outperforms the state-of-the-art model in terms of performance. Further ablation experiments also demonstrate the effectiveness of key components of the Disen-CGCN model, including disentangled representation learning, attention mechanisms, and personalized feature transformation. We also do a visual analysis of users' different preferences for items in different behaviors, demonstrating that our Disen-CGCN can capture more fine-grained user preferences in different behaviors.

In this study, we introduced the Disen-CGCN model, an advanced multi-behavior recommendation framework  that effectively utilizes the cascading relationships between different user behaviors. Our model distinctively incorporates disentangled representation techniques and attention mechanisms within each behavior to accurately discern diverse user preferences for items. Additionally, we integrate a meta-network between behaviors to facilitate personalized feature transformation for both users and items. Extensive experiments on two real datasets demonstrate that the Disen-CGCN model outperforms the state-of-the-art model in terms of accuracy with a large margin. Further ablation experiments also demonstrate the effectiveness of key components of the Disen-CGCN model, including disentangled representation learning, attention mechanisms, and personalized feature transformation. Moreover, through visual analysis, we provide insights into how Disen-CGCN effectively captures the intricate and varied user preferences across different behaviors, confirming its ability to provide more detailed and user-specific recommendations.

%\bibliographystyle{unsrt}

\normalem
\bibliographystyle{ACM-Reference-Format}
\bibliography{main}
%\bibliography{sample-base}

\end{document}